\documentclass[a4paper,twocolumn,11pt,accepted=2026-04-22]{quantumarticle}
\pdfoutput=1
\usepackage[utf8]{inputenc}
\usepackage[english]{babel}
\usepackage[table]{xcolor}
\usepackage[T1]{fontenc}
\usepackage{amsmath}
\usepackage{amssymb}
\usepackage{hyperref}
\usepackage{tikz}
\usepackage{lipsum}
\usepackage{graphicx}
\usepackage{epsfig}
\usepackage{epstopdf}
\usepackage{braket}
\usepackage{hyperref}
\usepackage{amsthm}
\usepackage{enumitem}
\usepackage{dsfont}
\usepackage{bbold}
\usepackage{bm}
\usepackage{eurosym}
\usepackage{diagbox}
\usepackage[numbers,sort&compress]{natbib}
\usepackage[ruled,vlined]{algorithm2e}
\usepackage{comment}
\usepackage{xcolor}
\usepackage[newfloat,frozencache,cachedir=.]{minted}
\setminted{
  fontsize=\small,
  linenos,
  numbersep=5pt,
  breaklines,
  tabsize=2,
  frame=lines,
  framesep=2mm,
  bgcolor=gray!5
}

\newcommand{\be}{\begin{equation}}
\newcommand{\ee}{\end{equation}}
\newcommand{\ba}{\begin{aligned}}
\newcommand{\ea}{\end{aligned}}

\newcommand{\bc}{\begin{center}}
\newcommand{\ec}{\end{center}}
\newcommand{\beq}{\begin{equation}}
\newcommand{\eeq}{\end{equation}}
\newcommand{\beqq}{\begin{equation*}}
\newcommand{\eeqq}{\end{equation*}}
\newcommand{\beqa}{\begin{align}}
\newcommand{\eeqa}{\end{align}}
\newcommand{\barr}{\begin{array}}
\newcommand{\earr}{\end{array}}
\newcommand{\bi}{\begin{itemize}}
\newcommand{\ei}{\end{itemize}}

\newtheorem{lem}{Lemma}
\newtheorem{theo}{Theorem}

\newtheorem{defi}{Definition}

\DeclareMathOperator{\Tr}{Tr}

\begin{document}

\title{Assessing non-Gaussian quantum state conversion with the stellar rank}

\author{Oliver Hahn}
\orcid{0000-0003-1677-8696}
\affiliation{Department of Basic Science, The University of Tokyo, 3-8-1 Komaba, Meguro-ku, Tokyo, 153-8902, Japan}
\affiliation{Wallenberg Centre for Quantum Technology, Department of Microtechnology and Nanoscience, Chalmers University of Technology, Sweden , SE-412 96 G\"{o}teborg, Sweden}

\author{Maxime Garnier}
\affiliation{DIENS, \'Ecole Normale Sup\'erieure, PSL University, CNRS, INRIA, 45 rue d’Ulm, Paris, 75005, France}

\author{Giulia Ferrini}
\affiliation{Wallenberg Centre for Quantum Technology, Department of Microtechnology and Nanoscience, Chalmers University of Technology, Sweden , SE-412 96 G\"{o}teborg, Sweden}

\author{Alessandro Ferraro}
\affiliation{Dipartimento di Fisica ``Aldo Pontremoli'',
Università degli Studi di Milano, I-20133 Milano, Italy}
\affiliation{Centre for Theoretical Atomic, Molecular and Optical Physics, Queen's University Belfast, Belfast BT7 1NN, United Kingdom}

\author{Ulysse Chabaud}
\email{ulysse.chabaud@inria.fr}
\orcid{0000-0003-0135-9819}
\affiliation{DIENS, \'Ecole Normale Sup\'erieure, PSL University, CNRS, INRIA, 45 rue d’Ulm, Paris, 75005, France}

\maketitle


\begin{abstract}
State conversion is a fundamental task in quantum information processing. Quantum resource theories allow for analyzing and bounding conversions that use restricted sets of operations. In the context of continuous-variable systems, state conversions restricted to Gaussian operations are crucial for both fundamental and practical reasons, particularly in state preparation and quantum computing with bosonic codes. However, previous analysis did not consider the relevant case of approximate state conversion. In this work, we introduce a framework for assessing approximate Gaussian state conversion by extending the stellar rank to the approximate stellar rank, which serves as an operational measure of non-Gaussianity. We derive bounds for Gaussian state conversion and distillation under approximate and probabilistic conditions, yielding new no-go results for non-Gaussian state preparation and enabling a reliable assessment of the performance of Gaussian conversion protocols. We also provide an open-source Python library to compute stellar-rank-related quantities and to assess Gaussian conversion.
\end{abstract}


\section{Introduction}
    
Understanding the boundaries between quantum computing architectures that are capable of providing quantum advantage for computation from those which are not is of crucial importance for the design of useful quantum computers. A tool to develop this understanding is the use of resource theories \cite{RevModPhys.91.025001}. In this framework, a set of states, operations and measurements is designed to be “free”. Often, this notion corresponds to the ease in the experimental capability of implementing such circuits elements. For notable examples, theorems have been derived that restrict the power of architectures made of solely free states, operations and measurements. 

Quantum computers over continuous-variable (CV) systems \cite{lloyd1999quantum, PhysRevLett.97.110501, RevModPhys.84.621, pfister2019continuous, fukui_building_2022-1},  where information is encoded via bosonic codes \cite{terhal_towards_2020, grimsmo_quantum_2021, joshi2021quantum, cai2021bosonic, albert2022bosonic, brady2024advances}, have emerged as an alternative to the ones based on two-level systems. In the context of these systems, a paradigmatic divide between “free” and “resourceful” circuit elements  is the one between Gaussian \cite{ferraro2005, adesso2014continuous} and non-Gaussian \cite{walschaers2021non} circuit elements. Gaussian operations are generally regarded as easier to achieve experimentally, in particular with optical set-ups \cite{RevModPhys.84.621,adesso2014continuous}. At the same time, circuits that are solely based on Gaussian elements are classically efficiently simulatable, i.e., they cannot provide exponential speed up for computation~\cite{Bartlett2002}. This result has also been extended to incorporate further simulatable architectures based on quasi-probability distributions; for instance circuits where the Wigner function is  non-negative everywhere are simulatable~\cite{mari2012positive,Veitch_2013,PhysRevLett.88.097904, PhysRevX.6.021039}. In this scenario, non-Gaussian states emerge as necessary resources for implementing universal quantum computation based on CV systems~\cite{walschaers2021non}.

The framework of resource theories is  instrumental to characterize the convertibility between resourceful states by means of free operations. It is indeed still an open question which non-Gaussian states actually provide quantum advantage for computation when supplied to Gaussian (simulatable) architectures. Therefore, showing the convertibility of non-Gaussian states to known resource states that unlock computational universality is a relevant way to assess the actual resourcefulness of general non-Gaussian states. Moreover, if some non-Gaussian quantum states can be prepared experimentally and Gaussian operations are easier to achieve, Gaussian conversion may unlock the ability of preparing new non-Gaussian quantum states in the lab; hence, characterizing convertibility of non-Gaussian states is also useful for quantum state preparation.
Such characterization can be achieved by identifying suitable measures of resourcefulness, i.e., monotones quantifying the amount of resourcefulness of a state, and by deriving bounds that constrain the convertibility of states with different amounts of resourcefulness.

Various measures and indicators of non-Gaussianity have been proposed so far~\cite{walschaers2021non}. These include negativity of the Wigner function, or Wigner negativity for short~\cite{kenfack2004negativity, albarelli2018resource,PhysRevA.97.062337}, stellar rank~\cite{chabaud2020stellar,chabaud2021holomorphic} or $n$-photon genuine quantum non-Gaussianity~\cite{Lachman2019faithful},  Gaussian rank and Gaussian extent~\cite{hahn2024classicalsimulationquantumresource, dias2024classical}. For the case of Wigner negativity, a resource theory has been developed, which allows for deriving bounds restricting which exact state conversions are in principle possible~\cite{albarelli2018resource,PhysRevA.97.062337}. 
However, all the bounds that were derived so far do not take into account the experimentally relevant case of approximate state conversion, where we wish to convert a resourceful state into another resourceful state approximately, i.e., up to a certain precision in suitable measures of similarity such as fidelity or trace distance.

In this work, we close this gap and introduce a general framework for assessing Gaussian state conversion in the approximate setting. To achieve this, we generalise the stellar rank \cite{chabaud2020stellar,chabaud2021holomorphic} and introduce the notion of $\epsilon$-approximate stellar rank of a state $\bm\rho$, as the minimum stellar rank of states that are $\epsilon$-close to $\bm\rho$ in fidelity. We demonstrate that the $\epsilon$-approximate stellar rank is a valid measure of non-Gaussianity, i.e., that it satisfies the properties required for being a monotone. Based on this new measure of non-Gaussianity, for a given precision $\delta\ge0$, we provide an $\epsilon$-parameterized family of bounds that, when simultaneously satisfied, indicate potential convertibility between non-Gaussian states, while when violated for any single value of the parameter $\epsilon$ rule out the possibility of approximate Gaussian convertibility of the input to the target state within trace distance less or equal to $\delta$. 
Moreover, we also show that the bound for $\epsilon=0$ also constrains probabilistic Gaussian protocols, in which one can post-select on specific outcomes of Gaussian measurements. As a byproduct we obtain that the stellar rank itself is a monotone under post-selected Gaussian protocols.

Based on these conversion bounds, we assess the efficiency of Gaussian conversion scenarios inspired by protocols that were previously studied in the literature. 
Importantly, we show that the $\epsilon$-approximate stellar rank can be computed efficiently in most relevant cases. 
Our framework allows to study and meaningfully assess Gaussian conversion and in particular non-Gaussian state breeding protocols, even when the input states are mixed, which are instrumental for non-Gaussian quantum state preparation and quantum computing with bosonic systems.

Additionally, we provide a robust, modular and open-source Python library to evaluate these bounds numerically \cite{stellarnumerics}. This software enables the computation of so-called stellar fidelities \cite{chabaud2020certification} thus making it useful for other applications beyond assessing Gaussian convertibility.

The paper is structured as follows. In section~\ref{sec:background}, we provide some background on resource theories and continuous-variable quantum information theory, including the stellar rank. In section~\ref{sec:approxsr}, we introduce formally the approximate stellar rank, demonstrate its main properties and explain how it can be computed. In section~\ref{sec:conversion}, we use this new non-Gaussianity measure to derive two kinds of Gaussian conversion bounds, the first one applying to deterministic, exact and approximate Gaussian protocols, and the second one applying to probabilistic, exact and approximate Gaussian protocols; we further use these bounds to obtain new no-go results for Gaussian conversion. In section~\ref{sec:num}, we present applications of these bounds for assessing the quality of Gaussian conversion protocols. We conclude in section~\ref{sec:conclusion}.


\section{Background}
\label{sec:background}


\subsection{Resource theories}
\label{sec:resource}

\noindent In this section we provide a brief review of resource theories and refer the reader to~\cite{albarelli2018resource,RevModPhys.91.025001} for further details.

A resource theory is characterized by a set of free states $\mathcal{G}$ and a set of free operations $\mathcal{F}$. The set of free states is closed under the action of free operations, namely it holds for $ \forall g\in\mathcal{G}$ that $\Lambda(g)\in\mathcal{G}$, $\forall\Lambda\in\mathcal{F}$. Every state not contained in the set of free states $\mathcal{G}$ is called a $\textit{resource}$. Resource monotones have been introduced to quantifiy the resource content of a state. 

\begin{defi}[Resource monotone]
A mapping $\mathcal{M}$ from the set of all states to the real numbers is called a resource monotone, if it is non-increasing under the set of free operations $\mathcal{F}$, i.e., for all density operator $\rho$,
\begin{align}\label{eq:conv-bounddet}
    \mathcal{M}(\rho) \geq  \mathcal{M}(\Lambda (\rho)), \;\; \forall \Lambda \in \mathcal{F}.
\end{align}
\end{defi}

\noindent Monotones in the literature do not necessarily fulfill all properties listed below, but they are useful in their own right and are used in this manuscript.

\begin{defi}[Faithfulness]
A monotone is called faithful if there is a constant $c \in \mathds{R}$ such that $\mathcal{M}(\rho)=c,\, \forall \rho\in\mathcal{G}$ and $\mathcal{M}(\rho)>c$ otherwise.
\end{defi}

\noindent The most common conventions are either $c=0$ if the monotone is (sub-)additive and $c=1$ if the monotone is (sub-)multiplicative.
 
\begin{defi}[Additivity and sub-additivity]
A monotone $\mathcal{M}$ is called additive if 
\begin{align}
    \mathcal{M}(\rho \otimes \sigma) = \mathcal{M}(\rho) +  \mathcal{M}( \sigma).
\end{align}
A monotone $\mathcal{M}$ is called sub-additive if 
\begin{align}
    \mathcal{M}(\rho \otimes \sigma) \leq \mathcal{M}(\rho) +  \mathcal{M}( \sigma).
\end{align}
\end{defi}

\noindent (Sub-)additivity is a very useful property, since it ensure that preparing a free state in an auxiliary system does not increase the amount of resource.

A strong property a monotone can satisfy is monotonicity under post-selection, i.e., the ability to post-select on a specific outcome of a measurement:

\begin{defi}[Monotonicity under post-selection]
A monotone $\mathcal{M}$ is called monotone under post-selection if for any free positive operator-valued measure $\{\Pi_{\bm\lambda}\}_{\bm\lambda}$
\begin{align}
    \mathcal{M}(\rho) \geq\mathcal{M}(\rho_{\bm{\lambda}}),
\end{align}
for $\rho_{\bm{\lambda}}= \frac{1}{\Tr(\Pi_{\bm{\lambda}}\rho)} \Pi_{\bm{\lambda}}\rho \Pi_{\bm{\lambda}}$.
\end{defi}

\noindent A monotone having the property of monotonicity under post-selection allows for computing bounds including probabilistic protocols. Indeed, given a resource monotone $\mathcal M$, if $\Lambda$ is a free operation that maps $k$ copies of $\rho$ to $m$ copies of $\sigma$ with any nonzero probability, then monotonicity under post-selection implies:
\begin{equation}\label{eq:conv-bound}
    \mathcal M(\rho^{\otimes k})\ge\mathcal M(\sigma^{\otimes m}).
\end{equation}


\subsection{Continuous-variable quantum information}
\label{sec:CV}

In this section we provide a brief review of CV quantum information theory, with particular emphasis on Gaussian quantum information in section \ref{sec:Gaussian} and the stellar representation of non-Gaussian states in section \ref{sec:stellar}. We refer the reader to~\cite{RevModPhys.84.621,chabaud2020stellar, chabaud2021holomorphic} for more in-depth treatments.

\subsubsection{Gaussian quantum information theory}
\label{sec:Gaussian}

\noindent Gaussian quantum optics offers a rich area of research, where a vast set of tools and analytical techniques are available~\cite{RevModPhys.84.621,ferraro2005}.
In this work, we denote the creation and annihilation operators by $\hat{a},\hat{a}^\dagger$, which satisfy the canonical commutation relation
\begin{align}
    [\hat{a},\hat a^\dag]=\hat{\mathbb 1}.
\end{align}
We denote by $\bm{\hat a}=(\hat a_1,\dots,\hat a_m)^T$ the tuple of annihilation operators over $m$ modes.
Gaussian unitaries are defined as 
\begin{align}
    \hat U_G&=e^{-i \hat H_G},
\end{align}
where $\hat H_G$ is at most quadratic in the creation and annihilation operators of the modes:
\begin{align}
    \hat{H}_G=\bm{\alpha}^T\bm{\hat{a}^\dagger}+\bm{\hat{a}^\dagger}U \bm{\hat{a}}+\bm{\hat{a}^\dagger} S \bm{\hat{a}^\dagger}+h.c.,
\end{align}
for $\bm{\alpha}\in\mathds{C}^m$ and where $U,S$ are $m\times m$ complex matrices for $m$ modes.
Gaussian unitary operators map creation and annihilation operators into affine symplectic combinations of creation and annihilation operators.

A generic single-mode Gaussian unitary can be decomposed into a sequence of displacements, squeezing and phase shifts:
\begin{align}
    \hat{D}(\alpha)&=e^{\alpha\hat{a}^\dag -\alpha^* \hat{a}},\\
    \hat{S}(\xi) &= e^{\frac{1}{2}(\xi \hat{a}^{\dag2}-\xi^* \hat{a}^2)},\\
    \hat{R}(\phi)&=e^{i \phi\,\hat{a}^\dagger \hat{a}}.
\end{align}

To decompose multimode Gaussian unitaries, an additional non-trivial $2$-mode unitary is sufficient. For example, one can consider the operation associated to the action of a beam splitter:
\begin{align}
    \hat{U}_{BS}(\theta)=e^{\theta(\hat{a}^\dagger_1\hat{a}_2-\hat{a}_1\hat{a}^\dagger_2)},
\end{align}
where the variable $\theta$ sets the transmissivity of the beamsplitter $\tau = \cos(\theta)^2$ --- which is called balanced if $\tau = \frac{1}{2}$ or equivalently $\theta = \frac{\pi}{4}$.

More generally, one can define the passive Gaussian operations as those that conserve the total number of photons. These unitary operators are generated by Hamiltonians of the form $\hat{H}_p=\bm{\hat{a}^\dagger}U\bm{\hat{a}} +h.c.$.
Any (multimode) Gaussian unitary can, as a consequence of the Euler (sometimes called Bloch--Messiah) decomposition, be decomposed as
\begin{align}
    \hat U_G= \hat{U} \hat{S}(\bm{\xi}) \hat{D}(\bm{\alpha})   \hat{V},
\end{align}
where $\hat{U}$ and $\hat{V}$ are passive Gaussian unitary operators.

A general Gaussian state is defined as 
\begin{align}
    \rho_G=\frac{e^{-\beta \hat H_G}}{\Tr[e^{-\beta\hat H_G}]},
\end{align}
including the case $\beta\rightarrow\infty$ for pure states. For pure states, this is equivalent to
\begin{align}
    \ket{\psi_G}&=\hat U_G \ket{\bm0}\\
    &= \hat{U} \hat{S}(\bm{\xi}) \hat{D}(\bm{\alpha})\ket{\bm0},
\end{align}
where $\ket{\bm0}$ denotes the tensor product of $m$ vacua and where we used the fact that passive Gaussian unitary operators map the vacuum state to itself in the last line.

We define Gaussian protocols as follows~\cite{albarelli2018resource}:

\begin{defi}[Gaussian protocol]\label{def:G_prot}
    A Gaussian protocol is a completely positive trace-preserving map composed of the following operations
    \begin{itemize}
        \item Gaussian unitaries: $\rho \mapsto \hat{U}_G \rho \hat{U}_G^\dagger$
        \item Composition with pure Gaussian state: $\rho \mapsto \rho \otimes \ket{\psi_G}\!\bra{\psi_G}$.
        \item Partial trace on subsystems: $\rho \mapsto \Tr_S[\rho]$.
        \item The above operations conditioned on classical randomness or Gaussian measurement outcomes.
    \end{itemize}
\end{defi}

\noindent This definition encompasses all Gaussian protocols based on quantum channels. However, probabilistic operations based on post-selection are also an important part of quantum state engineering, especially when one is considering quantum optics where genuine non-Gaussianity is typically generated by heralding specific measurement outcomes.
To that end, we define post-selected Gaussian protocols as follows~\cite{albarelli2018resource}:

\begin{defi}[Post-selected Gaussian protocol]\label{def:G_prot2}
    A post-selected Gaussian protocol is a linear map composed of the following operations
    \begin{itemize}
        \item Gaussian unitaries: $\rho \mapsto \hat{U}_G \rho \hat{U}_G^\dagger$
        \item Composition with pure Gaussian state: $\rho \mapsto \rho \otimes \ket{\psi_G}\!\bra{\psi_G}$.
        \item Partial trace on subsystems: $\rho \mapsto \Tr_S[\rho]$.
        \item Post-selected pure Gaussian measurement of subsystem: $\rho \mapsto \Tr_S[\rho (\mathds{1}\otimes \ket{\psi_G(\bm{\alpha})}\bra{\psi_G(\bm{\alpha})} )]/ p(\bm{\alpha}|\rho)$ with $p(\bm{\alpha}|\rho)\!=\! \Tr[  \rho (\mathds{1}\otimes \ket{\psi_G(\bm{\alpha})}\bra{\psi_G(\bm{\alpha})}) ]$ and where $\bm{\alpha}$ is a vector of measurement outcomes.
        \item The above operations conditioned on classical randomness or Gaussian measurement outcomes.
    \end{itemize}
\end{defi}

Despite the richness of Gaussian quantum optics, we cannot represent all possible states and operations using the Gaussian manifold. More critically, when restricting to Gaussian protocols acting on Gaussian states, achieving exponential quantum computational advantage becomes impossible, as such processes can be efficiently simulated using classical algorithms~\cite{RevModPhys.84.621}. The setting of Gaussian quantum information processing is moreover limited by several no-go results for various tasks, including entanglement distillation~\cite{PhysRevLett.89.097901,PhysRevLett.89.137904,PhysRevA.66.032316}, error correction~\cite{PhysRevLett.102.120501}, and violation of Bell inequalities/contextuality~\cite{PhysRevLett.129.230401}. Thus, in order to obtain genuine quantum phenomena beyond what is classical simulatable, we need to include non-Gaussian operations or states.

\subsubsection{The stellar representation of non-Gaussian states}
\label{sec:stellar}

The stellar representation has been introduced in~\cite{chabaud2020stellar} to characterize single-mode non-Gaussian quantum states and is related to the notion of genuine $n$-photon non-Gaussianity \cite{Lachman2019faithful}. It has later been generalised to the multimode setting in~\cite{chabaud2020classical,chabaud2021holomorphic,yao2024riemannian}. In this section, we give a brief review of this formalism.

The stellar representation classifies CV quantum states using the so-called \textit{stellar function} (or Bargmann, Fock--Bargmann, Segal--Bargmann function), which is a representation of a quantum state in an infinite-dimensional Hilbert space by a holomorphic function originally due to Segal~\cite{segal1963mathematical} and Bargmann~\cite{bargmann1961hilbert}.
More precisely, to any $m$-mode pure quantum state $\ket{\bm\psi}=\sum_{\bm n\ge\bm 0}\psi_{\bm n}\ket{\bm n}$ is associated its stellar function
\begin{equation}
    F^\star_{\bm\psi}(\bm z):=\sum_{\bm n\ge\bm 0}\frac{\psi_{\bm n}}{\sqrt{\bm n!}}\bm z^{\bm n},
\end{equation}
for all $\bm z\in\mathbb C^m$.

The algebraic structure of this function allows us to rank CV quantum states using the \textit{stellar rank}: pure quantum states of finite stellar rank $r^\star\in\mathbb N$ are those states whose stellar function is of the form $P\times G$, where $P$ is a multivariate polynomial of degree $r^\star$ and $G$ is a multivariate Gaussian function, while the other states are of infinite stellar rank.
For a mixed state $\bm\rho$, the stellar rank is defined using a convex roof construction: ${r^\star(\bm\rho)=\inf_{p_i,\bm\psi_i}\sup r^\star_{\bm\psi_i}}$, where the infinimum is over the decompositions $\bm\rho={\sum_ip_i\ket{\bm\psi_i}\!\bra{\bm\psi_i}}$.

The stellar rank possesses remarkable properties---below we recall the ones which are used hereafter. 

\begin{figure*}[t]
\centering
\includegraphics[width=1\linewidth]{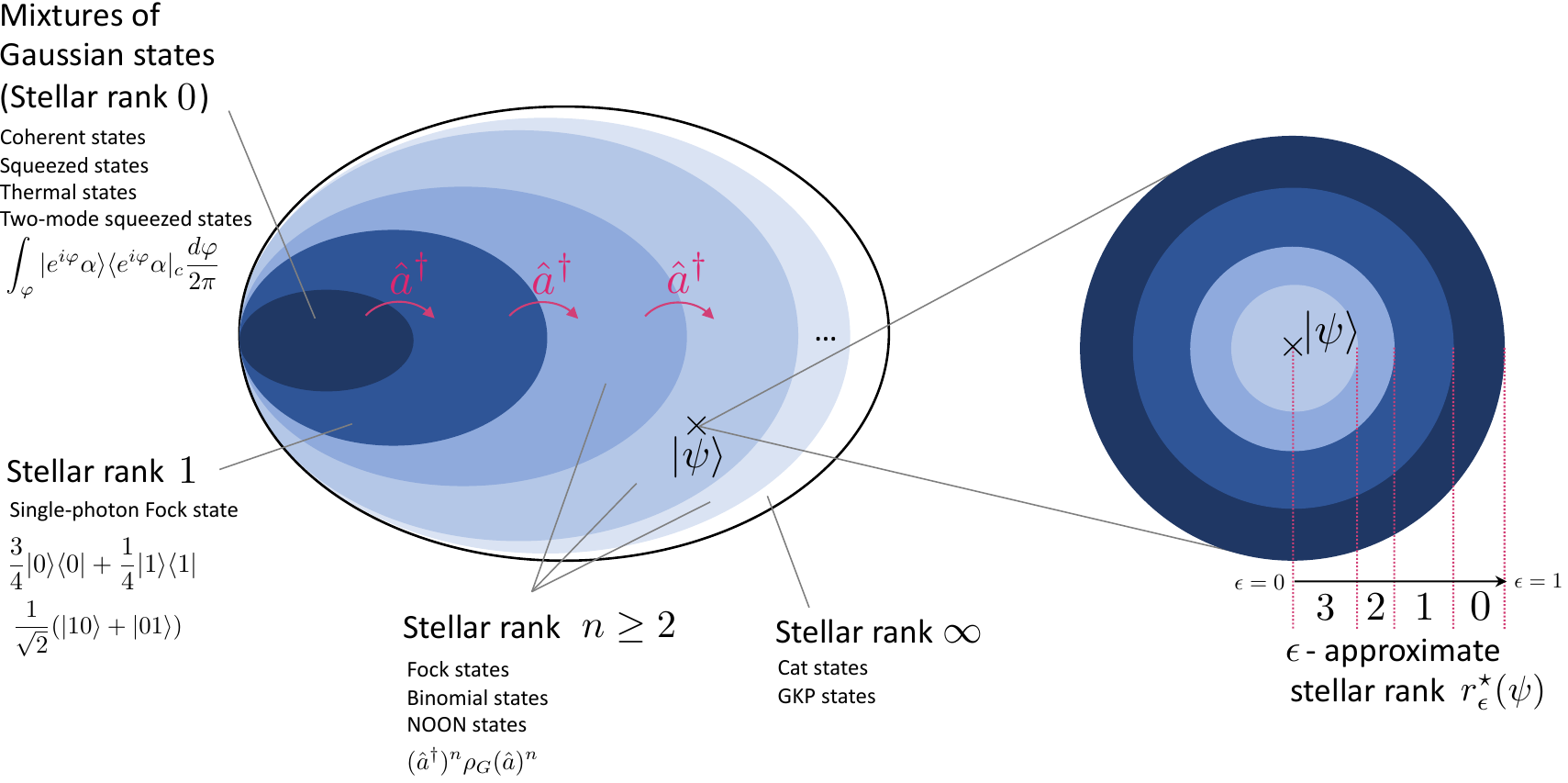}
\caption{The set of quantum states, partitioned following the multimode stellar hierarchy (extended from \cite{chabaud2021holomorphic}). For each rank, we give examples of states that are commonly encountered in the literature. The last example of stellar rank $0$ is a rotationally-invariant mixture of coherent states. The density matrix $\rho_G$ denotes a Gaussian state. Gaussian unitary operators leave the stellar rank invariant, while non-Gaussian operators can increase it (such as the creation operator, which increases the stellar rank by one). The right hand side is a schematic representation of the topology of the hierarchy, at the vicinity of a pure state $\ket\psi$ of stellar rank $3$, illustrating how the $\epsilon$-approximate stellar rank decreases the further away one gets from $\ket\psi$.}
\label{fig:s-hier}
\end{figure*}

A unitary  operation  is  Gaussian  if  and  only  if  it  leaves  the stellar  rank invariant, and the stellar rank is non-increasing under Gaussian channels and measurements \cite{chabaud2021holomorphic}. As a result, the stellar rank is a non-Gaussian monotone which provides a measure of the non-Gaussian character of a quantum state and induces a non-Gaussian hierarchy among quantum states, the \textit{stellar hierarchy} (see Fig.\ \ref{fig:s-hier}). At rank $0$ of the hierarchy lie mixtures of Gaussian states, while quantum non-Gaussian states \cite{filip2011detecting} populate all higher ranks. For instance, a multimode Fock state $\ket{\bm n}$ has stellar rank $|\bm n|$ \cite{chabaud2021holomorphic}, while a single-mode cat state has infinite stellar rank \cite{chabaud2020stellar}.

The stellar hierarchy bears an operational significance, because the stellar function gives a prescription for engineering a pure quantum state from the vacuum: $\ket{\bm\psi}=F^\star_{\bm\psi}(\hat a^\dag_1,\dots,\hat a^\dag_m)\ket{\bm0}$. As a result, a state of stellar rank $n$ cannot be obtained from the vacuum by using less than $n$ applications of creation operators, together with Gaussian unitary operations. In other words, the stellar rank has an operational interpretation as a lower bound on the minimal number of elementary non-Gaussian operations (applications of a single-mode creation operator, i.e.\ single-photon addition) needed to engineer the state from the vacuum.

The stellar hierarchy is robust with respect to the trace distance, i.e.\ every state of a given finite stellar rank only has states of equal or higher rank in a close vicinity, whose size depends on the state \cite{chabaud2020stellar,chabaud2021holomorphic}. 
Moreover, states of finite stellar rank form a dense subset of the Hilbert space, so that states of infinite rank can be approximated arbitrarily well in trace distance by sequences of states of finite stellar rank. The optimal approximation for any fixed rank can be obtained by an optimisation over ${\cal O}(m^2)$ parameters for $m$-mode states, independently of the rank. As a consequence, the stellar rank of quantum states can be witnessed experimentally \cite{chabaud2020certification,fiuravsek2022efficient}. We elaborate on this property in section \ref{sec:comp-asr}.

Finally, CV quantum computations with low stellar rank can be simulated efficiently using a classical computer \cite{chabaud2020classical,chabaud2023resources}, which implies that the stellar rank is a necessary non-Gaussian resource for quantum computational speedup with CV.

Although it is a non-Gaussian monotone with valuable operational properties, the stellar rank is too crude of a measure when assessing Gaussian conversion protocols beyond exact conversion. Indeed, states of high stellar rank may be very close to states of much lower stellar rank, as can be illustrated by the states of the form $|\psi_{\epsilon,N}\rangle=\sqrt{1-\epsilon}\ket0+\sqrt\epsilon\ket N$ for small $\epsilon$ and large $N$. These are states of arbitrarily high stellar rank $N$, but up to a small perturbation $\epsilon$ they are effectively of stellar rank $0$. 

In what follows, we introduce a fine-grained measure based on the stellar rank that allows us to assess Gaussian conversion protocols in practical situations, i.e., in both the exact and approximate cases.


\section{The approximate stellar rank}
\label{sec:approxsr}

\subsection{Definition} 

In the context of CV quantum information processing, identifying the minimal number of non-Gaussian operations required to engineer a desired quantum state up to a given precision is of paramount importance, especially for quantum optics platforms where such operations are hard to implement in a deterministic fashion. 
For this purpose, we introduce a new measure based on the stellar rank: 

\begin{defi}[Approximate stellar rank]\label{def:asr}
For $0\le\epsilon\le1$, the $\epsilon$-approximate stellar rank of an $m$-mode density operator $\bm\rho$ is defined as
\begin{equation}\label{eq:def-asr}
    r^\star_\epsilon(\bm\rho):=\inf_{\bm\tau,F(\bm\rho,\bm\tau)\ge1-\epsilon}r^\star(\bm\tau),
\end{equation}
where $F(\bm\rho,\bm\tau)=\mathrm{Tr}\left(\sqrt{\sqrt{\bm\rho}\bm\tau\sqrt{\bm\rho}}\right)^{
\!2}$ denotes the fidelity. 
\end{defi}

\noindent The $\epsilon$-approximate stellar rank $r^\star_\epsilon$ is a generalisation of the stellar rank parametrised by $\epsilon\in[0,1]$, with $r^\star_0(\bm\rho)=r^\star(\bm\rho)$. Hereafter, we use the convention $r^\star_\epsilon=0$ for $\epsilon>1$. The infimum in the definition is defined with respect to the topology induced by the trace norm, and is actually a minimum, as the $\epsilon$-approximate stellar rank is an integer-valued, non-increasing function of $\epsilon$, with $r^\star_1(\bm\rho)=0$. It gives the minimal stellar rank of states that are $\epsilon$-close in fidelity to a given state. From a mathematical standpoint, it quantifies how well a stellar function may be approximated by a function that is the product of a polynomial and a Gaussian, for increasing degree of the polynomial. Given the operational properties of the stellar rank, the $\epsilon$-approximate stellar rank can thus be understood as a minimal cost in terms of non-Gaussian operations to prepare an approximation of a target state. 

A similar quantity known as the $\epsilon$-smoothed non-Gaussianity of formation as been previously introduced in the single-mode setting~\cite{chabaud2020stellar} and is readily generalised to the multimode setting: $\mathcal{NGF}_\epsilon(\bm\rho):=\min_{\bm\tau,D(\bm\rho,\bm\tau)\le\epsilon}r^\star(\bm\tau)$, where $D$ denotes the trace distance. By the Fuchs--van de Graaf inequalities, it satisfies $r^\star_{\epsilon}(\bm\psi)\le\mathcal{NGF}_\epsilon(\bm\psi)\le r^\star_{\epsilon^2}(\bm\psi)$ for pure states and $r^\star_{2\epsilon}(\bm\rho)\le\mathcal{NGF}_\epsilon(\bm\rho)\le r^\star_{\epsilon^2}(\bm\rho)$ in general, so these two quantities are operationally equivalent. Compared to the $\epsilon$-approximate stellar rank, the $\epsilon$-smoothed non-Gaussianity of formation measures the quality of the approximation using trace distance rather than fidelity. As a result, the $\epsilon$-smoothed non-Gaussianity of formation is challenging to compute, while the $\epsilon$-approximate stellar rank can be computed efficiently in several practical scenarios, as we show in section~\ref{sec:comp-asr}.

\subsection{Properties} 

The stellar rank is sub-additive \cite{chabaud2021holomorphic}. Let $\bm\rho$ and $\bm\sigma$ be two density operators, then
\begin{equation}\label{eq:subadd}
    \max[r^\star(\bm\rho),r^\star(\bm\sigma)]\le r^\star(\bm\rho\otimes\bm\sigma)\le r^\star(\bm\rho)+r^\star(\bm\sigma).
\end{equation}
In particular, it is additive if one of the density operators is a pure state or a mixture of Gaussian states.

On the other hand, for the $\epsilon$-approximate stellar rank, there are cases where $r^\star_\epsilon(\bm\rho\otimes\bm\sigma)> r^\star_\epsilon(\bm\rho)+r^\star_\epsilon(\bm\sigma)$,  even for pure states. For instance, consider the Fock state $\ket{1}$: writing $f:=3\sqrt3/(4e)\approx0.48$, we have $r^\star_\epsilon(\ket1)=1$ for $0\le\epsilon<1-f$ and $r^\star_\epsilon(\ket1)=0$ for $1-f\le\epsilon\le1$ \cite{filip2011detecting,chabaud2020stellar}, while $r^\star_\epsilon(\ket1^{\otimes2})=2$ for $0\le\epsilon<1-f$, $r^\star_\epsilon(\ket1^{\otimes2})=1$ for $1-f\le\epsilon<3/4$, and $r^\star_\epsilon(\ket1^{\otimes2})=0$ for $3/4\le\epsilon\le1$ \cite{chabaud2020certification,chabaud2025erratum}.
In particular, the $\epsilon$-approximate stellar rank of the Fock state $\ket1$ satisfies $1=r^\star_\epsilon(\ket1^{\otimes2})>2r^\star_\epsilon(\ket1)=0$ for $1-f\le\epsilon<3/4$. This means that for a fixed $\epsilon$, the $\epsilon$-approximate stellar rank is generally not sub-additive.
However, by allowing different values of $\epsilon$, we show that the approximate stellar rank does satisfy a sub-additivity property:

\begin{lem}[Sub-additivity of the approximate stellar rank]\label{lem:upperbound}
Let $k\ge1$ and let $\bm\rho_1,\dots,\bm\rho_k$ be (mixed) states. Let $\bm\epsilon=(\epsilon_1,\dots,\epsilon_k)\in[0,1]^k$ with $|\bm\epsilon|=\epsilon_1+\dots+\epsilon_k\le1$. Then
\begin{equation}
    \max_{i=1\dots k}r^\star_{|\bm\epsilon|}(\bm\rho_i)\le r^\star_{|\bm\epsilon|}\!\left(\bigotimes_{i=1}^k\bm\rho_i\right)\le\sum_{i=1}^kr^\star_{\epsilon_i}(\bm\rho_i).
\end{equation}
In particular, setting $\bm\rho_1=\dots=\bm\rho_k=\bm\rho$ and $\epsilon_1=\dots=\epsilon_k=\frac\epsilon k$ gives
\begin{equation}\label{eq:subaddeps}
    r^\star_\epsilon(\bm\rho)\le r^\star_\epsilon(\bm\rho^{\otimes k})\le k\,r^\star_{\epsilon/k}(\bm\rho).
\end{equation}
\end{lem}

\noindent We give a proof in Appendix~\ref{app:asr-nGmon}.

The relevance of the $\epsilon$-approximate stellar rank for Gaussian conversion comes from the following result:

\begin{theo}[The approximate stellar rank is a monotone]\label{th:asr-nGmon} For all $0\le\epsilon\le1$, the $\epsilon$-approximate stellar rank is a monotone under Gaussian protocols which vanishes for Gaussian states. Moreover, for $\epsilon=0$, the stellar rank is a monotone under post-selected Gaussian protocols.
\end{theo}

\noindent We give a proof in Appendix~\ref{app:asr-nGmon}. This result has important consequences for Gaussian conversion, which we detail in section~\ref{sec:conversion}.

\subsection{Computation} 
\label{sec:comp-asr}

The $\epsilon$-approximate stellar rank is closely related to the notion of maximum achievable fidelity using states of bounded stellar rank introduced in~\cite{chabaud2020certification}, or \textit{stellar fidelities} for short:

\begin{defi}[Stellar fidelities]\label{def:sF}
The stellar fidelities of a state $\bm\rho$ are defined as
\begin{equation}\label{eq:stellarF}
    f_n^\star(\bm\rho):=\sup_{\bm\tau,r^\star(\bm\tau)\le n}F(\bm\rho,\bm\tau),
\end{equation}
for all $n\in\mathbb N$.
\end{defi}

\noindent The supremum in the definition is defined with respect to the topology induced by the trace norm. The stellar fidelity $f^\star_0(\bm\psi)$ is the maximal fidelity with a mixture of Gaussian states, while $f_n^\star(\bm\rho)=1$ for all $n\ge r^\star(\bm\rho)$.
From Definition~\ref{def:asr} and Definition~\ref{def:sF}, we obtain the following relation between stellar fidelities and approximate stellar rank:

\begin{lem}[Equivalence between stellar fidelities and approximate stellar rank]\label{lem:asr-sF}
For all $n\in\mathbb N$ and all $0\le\epsilon\le1$,
\begin{equation}
    f^\star_n(\bm\rho)\ge1-\epsilon\Leftrightarrow\forall\epsilon'>\epsilon,\,r^\star_{\epsilon'}(\bm\rho)\le n,
\end{equation}
or equivalently, for all $1\le n<r^\star(\bm\rho)$,
\begin{equation}
    \!r^\star_\epsilon(\bm\rho) =\!
    \begin{cases} 
    r^\star(\bm\rho) & \!\!\!\!\text{for } 
    \epsilon\!\in\![0,1-f^\star_{r^\star(\bm\rho)-1}(\bm\rho)),\\
    n & \!\!\!\!\text{for } 
    \epsilon\!\in\!(1-f^\star_n(\bm\rho),1-f^\star_{n-1}(\bm\rho)),\\
    0 & \!\!\!\!\text{for } \epsilon\!\in\!(1-f^\star_0(\bm\rho),1].
    \end{cases}
\end{equation}
For pure states, this can be refined as
\begin{equation}
    f^\star_n(\bm\psi)\ge1-\epsilon\Leftrightarrow r^\star_\epsilon(\bm\psi)\le n,
\end{equation}
or equivalently,
\begin{equation}
    \!r^\star_\epsilon(\bm\psi) =\!
    \begin{cases} 
    r^\star(\bm\psi) & \!\!\!\!\text{for } 
    \epsilon\!\in\![0,1-f^\star_{r^\star(\bm\psi)-1}(\bm\psi)),\\
    n & \!\!\!\!\text{for } 
    \epsilon\!\in\![1-f^\star_n(\bm\psi),1-f^\star_{n-1}(\bm\psi)),\\
    0 & \!\!\!\!\text{for } \epsilon\!\in\![1-f^\star_0(\bm\psi),1].
    \end{cases}
\end{equation}
\end{lem}

\noindent We give a proof in Appendix~\ref{app:equiv}, where we show that the second part of the Lemma is a consequence of the fact that the supremum in the definition of stellar fidelities is a maximum for pure states.  
In particular, for a pure state $\bm\psi$, the stellar fidelity $f^\star_0(\bm\psi)$ is the maximal fidelity with a pure Gaussian state. 
Note here the connection with the Gaussian extent~\cite{hahn2024classicalsimulationquantumresource} that is an alternative non-Gaussian measure. For some classes of states, such as for example Fock states, the Gaussian extent is the inverse of the maximal fidelity with a pure Gaussian state \cite{lami2021framework,hahn2024classicalsimulationquantumresource}.

\begin{figure}[!ht]
\centering
\includegraphics[width=.8\linewidth]{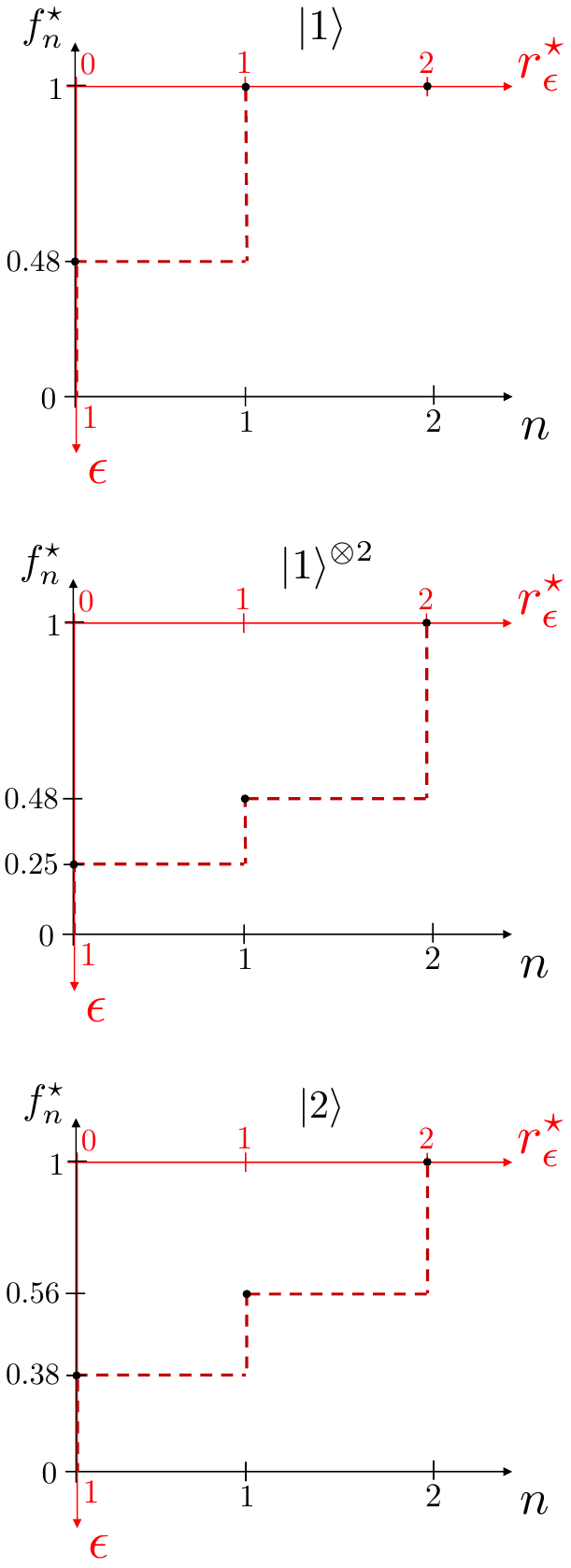}
\caption{Stellar fidelities (in black) and approximate stellar rank (in red) of the Fock states $\ket1$, $\ket1^{\otimes2}$ and $\ket2$, as a function of stellar rank $n$ and approximation parameter $\epsilon$, respectively. In each profile, each black dot represents a stellar fidelity $f_n^\star$, while each dashed line gives the $\epsilon$-approximate stellar rank.}
\label{fig:profile1and11}
\end{figure}

Lemma~\ref{lem:asr-sF} implies that the values of the stellar fidelities $f^\star_n(\bm\psi)$ for all $n$ specify the values of the $\epsilon$-approximate stellar rank $r^\star_\epsilon(\bm\psi)$ for all $0\le\epsilon\le1$, and vice versa, as illustrated in Fig.\ \ref{fig:profile1and11}.

As it turns out, computing the stellar fidelities of pure states over a few modes is efficient, because it can be reduced to an optimisation over a subset of Gaussian unitaries. Let us write, for all $n\in\mathbb N$,
\begin{equation}
    \bm\Pi_n=\sum_{|\bm p|\le n}\ket{\bm p}\!\bra{\bm p},
\end{equation}
the projector onto the subspace of $m$-mode states with at most $n$ photons. From Theorem 2 in~\cite{chabaud2020certification}:

\begin{theo}[Computation of stellar fidelities \cite{chabaud2020certification}]\label{th:profiles}
Let $n\in\mathbb N$ and let $\ket{\bm\psi}$ be a $m$-mode target pure state. Then, the maximum achievable fidelity with the target state $\ket{\bm\psi}$ using $m$-mode states of finite stellar rank less or equal to $n$ is given by
\begin{equation}\label{eq:profiles}
f_n^\star(\bm\psi)=\max_{\hat G}\bra{\bm\psi}\hat G^\dag\bm\Pi_n\hat G\ket{\bm\psi},
\end{equation}
where the supremum is over $m$-mode Gaussian unitary operations of the form $\hat G=\hat S\hat D\hat U$, where $\hat S$ is a tensor product of squeezing operators with real-valued squeezing parameters, $\hat D$ is a tensor product of displacement operators and $\hat U$ is a passive linear operator. Moreover, assuming the optimisation yields a Gaussian operation $\hat G_0$, an optimal approximating state of stellar rank at most $n$ is $\hat G_0^\dag\bm\Pi_n\hat G_0\ket{\bm\psi}$, up to normalisation.
\end{theo}

\noindent In particular, given a target state $\ket{\bm\psi}$, this result allows us to compute the set of achievable fidelities with that state using states of finite stellar rank $n$ for all $n\in\mathbb N$, i.e., the profile of stellar fidelities of $\ket{\bm\psi}$. The values of the $\epsilon$-approximate stellar rank for all $\epsilon$ can then be read from this profile, via Lemma~\ref{lem:asr-sF} (see Fig.~\ref{fig:profile1and11}). 

A limitation of this technique is that the numerical bounds derived do not come with a certificate: the stellar fidelities are obtained by an optimisation over a non-convex space, so we cannot be sure in general that the numerical value obtained corresponds to the global optimum. For many practical examples, however, the optimisation space is small enough so that we can have high confidence about the numerical bounds obtained. This is the case in particular for most single-mode examples, which lead to trustful bounds for multimode cases as well using the sub-additivity bounds from Lemma~\ref{lem:upperbound}, as we show in the next section. We provide a code library for computing stellar profiles in \cite{stellarnumerics} (see Appendix~\ref{app:num} for more details on the numerical implementation).

While stellar fidelities for pure states can be numerically computed via Theorem~\ref{th:profiles}, the case of mixed states is more challenging, because the closest state of bounded stellar rank can also be a mixed state. To address this issue, we define \textit{pure stellar fidelities} as
\begin{equation}\label{eq:puresf}
f_{
n,\mathrm{pure}}^\star(\bm\rho):=\max_{\hat G}[\mathrm{max}\,\mathrm{eig}\,(\bm\Pi_n\hat G^\dag\bm\rho\hat G\bm\Pi_n)],
\end{equation}
where the largest eigenvalue is maximized over all Gaussian unitaries.
We show in Appendix~\ref{app:puresf} that the stellar fidelities of a mixed state $\bm\rho$ are lower bounded by its pure stellar fidelities: 
\begin{equation}\label{eq:puresfbound}
f_{
n,\mathrm{pure}}^\star(\bm\rho)\le f_
n^\star(\bm\rho),
\end{equation}
for all $n\in\mathbb N$, with equality if the state is pure. Since pure stellar fidelities can be computed numerically in a similar fashion to Theorem~\ref{th:profiles}, this bound provides a tool to assess Gaussian conversion involving mixed states, as we detail in the next section.

We emphasize that the central quantity of interest is the $\epsilon$-approximate stellar rank, for which we establish all the necessary properties of a monotone. 
The (pure) stellar fidelities serve as auxiliary quantities used to bound the $\epsilon$-approximate stellar rank, and their relevance for state conversion arises through this connection. 
Moreover, the $\epsilon$-approximate stellar rank provides a single scalar measure that quantifies the resourcefulness of a quantum state, with strong operational interpretations inherited from the stellar rank. 
In contrast, the stellar fidelities constitute, in principle, an infinite hierarchy of parameters, making the $\epsilon$-approximate stellar rank a far more compact and convenient quantity to work with.


\section{Gaussian conversion}
\label{sec:conversion}

Gaussian conversion amounts to producing non-Gaussian states from other non-Gaussian states via Gaussian protocols (see Definition \ref{def:G_prot} and Definition \ref{def:G_prot2}). It includes protocols such as non-Gaussian distillation or non-Gaussian state breeding, among others \cite{Ourjoumtsev2007generation,Etesse:2014ty,Weigand2018generating, takase2024generation,zheng2020gaussian, hahn2022deterministic,Zheng2023gaussian,Etesse:14,Arzani2017polynomial,PhysRevA.97.062337}. 

Quantum state conversion can be exact or approximate, deterministic or probabilistic (post-selected/heralded).
In what follows, we study both deterministic and probabilistic cases under the same umbrella. Using approximate stellar rank as our main tool, we focus on exact Gaussian conversion in section \ref{sec:exact} and on approximate Gaussian conversion in section \ref{sec:approximate}. We further obtain strong no-go results for Gaussian conversion in section \ref{sec:nogo}.
We illustrate our results with numerical examples inspired from Gaussian conversion protocols which have been studied theoretically and/or demonstrated experimentally in section \ref{sec:num}.

\subsection{Exact Gaussian conversion bounds}
\label{sec:exact}

In this section, we illustrate the use of approximate stellar rank for assessing exact Gaussian conversion. Note that exact conversion is a particular case of approximate conversion (when the error parameter is set to $0$) which we analyse in the next section~\ref{sec:approximate}.

Recall from section~\ref{sec:resource} and Eq.~(\ref{eq:conv-bounddet}) that any non-Gaussian monotone $\mathcal M$ yields 
\begin{equation}\label{eq:conv-bound-2}
    \mathcal M(\rho)\ge\mathcal M(\sigma)
\end{equation}
when exact Gaussian conversion of $\rho$ to $\sigma$ is possible.

The Wigner Logarithmic Negativity (WLN) is defined as $W(\bm\rho)=\log\int|W_{\bm\rho}|$, where $W_{\bm\rho}$ is the Wigner function. It has been used as a non-Gaussian monotone to assess exact Gaussian conversion protocols in~\cite{albarelli2018resource}, both in deterministic and probabilistic cases. Since it is a non-Gaussian monotone \cite{chabaud2021holomorphic}, the stellar rank can also be used to assess exact Gaussian conversion protocols, in a way that is somewhat complementary to negativity of the Wigner function.
For instance, a cat state and a Fock state $\ket1$ can have the same WLN for a well-chosen cat state amplitude, but the former has infinite stellar rank while the latter has finite stellar rank (equal to $1$), so exact Gaussian conversion from the Fock state to the cat state is impossible, which Wigner negativity cannot conclude.
However, quantum states with the same stellar rank can have different Wigner negativity. For instance, the Wigner negativity of the state $\ket1\otimes\ket1$ is about $0.71$, while that of the state $\ket2$ is about $0.55$ \cite{kenfack2004negativity}, even though both states have stellar rank $2$. 
Moreover, unlike stellar rank, Wigner negativity is robust to small perturbations: consider the example pointed out earlier of the states $\ket1$ and $\ket{\psi_{\epsilon,1}}=\sqrt{1-\epsilon}\ket0+\sqrt\epsilon\ket1$; both states have stellar rank $1$, but Wigner negativity clearly tells us that conversion from the latter state to the former is impossible for small $\epsilon$.

Hereafter, we show that the $\epsilon$-approximate stellar rank provides a good compromise when assessing exact Gaussian conversion. For convenience in what follows, whenever we obtain bounds based on the approximate stellar rank, we provide an alternative expression based on stellar fidelities via Lemma~\ref{lem:asr-sF}. 

\begin{lem}[Exact conversion bounds]\label{lem:asr-conv-bound-exact}
If $\bm\sigma$ can be obtained from $\bm\rho$ using a Gaussian protocol, then for all $0\le\epsilon\le1$ and all $n\in\mathbb N$:
\begin{equation}\label{eq:asr-conv-bound-exact}
    \begin{cases}
        r^\star_\epsilon(\bm\rho)\ge r^\star_\epsilon(\bm\sigma),\\
        f^\star_n(\bm\rho)\le f^\star_n(\bm\sigma).
    \end{cases}
\end{equation}
Moreover, if $\bm\sigma$ can be obtained from $\bm\rho$ using a post-selected Gaussian protocol with any nonzero probability, then:
\begin{equation}\label{eq:asr-conv-bound-exact0}
    r^\star(\bm\rho)\ge r^\star(\bm\sigma).
\end{equation}
\end{lem}

\noindent We give a quick proof in Appendix \ref{app:sF-conv-bound-exact}.
From a graphical perspective, this result implies that the profile of stellar fidelities of the input has to be below that of the output if conversion via a Gaussian protocol is possible.

Let us illustrate the use of these bounds for the Gaussian conversion $\ket1\mapsto\ket2$. In this case, the WLNs are given by $W(\ket1)\simeq0.35$ and $W(\ket2)\simeq0.55$ \cite{kenfack2004negativity}. Since WLN is monotone on average \cite{albarelli2018resource}, we obtain $W(\ket1)\ge p\,W(\ket2)$ if exact Gaussian conversion is possible with probability $p$, which implies that exact Gaussian conversion from $\ket1$ to $\ket2$ with probability $p$ is impossible whenever the probability of success satisfies $p>0.35/0.55\simeq0.64$. In comparison, Eq.~(\ref{eq:asr-conv-bound-exact0}) implies that exact Gaussian conversion from $\ket1$ to $\ket2$ is actually impossible \textit{for any probability} $p$.

As this toy example illustrates, Lemma~\ref{lem:asr-conv-bound-exact} constitutes a powerful tool to rule out exact Gaussian conversion, leading to a large family of no-goes beyond state-of-the-art, as we show in section~\ref{sec:nogo}. More importantly, we generalise it to approximate Gaussian conversion in the next section.

\subsection{Approximate Gaussian conversion bounds}
\label{sec:approximate}

The bounds derived in the previous section can be extended to the more general case of approximate Gaussian conversion, in which the output state of the conversion protocol is only guaranteed to be $\delta$-close in trace distance to the target state, for some error parameter $\delta\ge0$:

\begin{theo}[Approximate conversion bounds]\label{th:asr-conv-bound-approx}
If a state $\delta$-close in trace distance to $\bm\sigma$ can be obtained from $\bm\rho$ using a Gaussian protocol, then for all $0\le\epsilon\le1$ and all $n\in\mathbb N$,
\begin{equation}\label{eq:asr-conv-bound-approximate}
    \begin{cases}
        r^\star_\epsilon(\bm\rho)\ge r^\star_{\epsilon+\sqrt{2\delta}}(\bm\sigma),\\
        f^\star_n(\bm\rho)\le f^\star_n(\bm\sigma)+\sqrt{2\delta}.
    \end{cases}
\end{equation}
Moreover, if a state $\delta$-close in trace distance to $\bm\sigma$ can be obtained from $\bm\rho$ using a post-selected Gaussian protocol with any nonzero probability, then:
\begin{equation}\label{eq:asr-conv-bound-approximate0}
    r^\star(\bm\rho)\ge r^\star_{\sqrt{2\delta}}(\bm\sigma).
\end{equation}
\end{theo}

\noindent We give a proof in Appendix~\ref{app:asr-conv-bound-approx}, where we also show that these bounds may be refined when $\bm\rho=\bm\psi$ and $\bm\sigma=\bm\phi$ are pure states by replacing $\sqrt{2\delta}$ by $\delta$.
These conditions imply that for each stellar rank the maximum achievable fidelity with $\bm\rho$ is greater than the maximum achievable fidelity with $\bm\sigma$ by at most an amount $\sqrt{2\delta}$ ($\delta$ for pure states). 
From a graphical perspective, this means that the profile of stellar fidelities of the input is below that of the output, after lowering the former by a vertical offset depending on $\delta$. 
Moreover, a violation of the bound in Eq.~(\ref{eq:asr-conv-bound-approximate0}) witnesses impossibility of approximate Gaussian conversion from ${\bm\rho}$ to ${\bm\sigma}$ with trace distance error smaller than $\delta$, even when allowing for post-selection \textit{for any nonzero probability of success}.

In order to use these theoretical bounds in practice, we need to compute the approximate stellar rank. For pure states over few modes, we can compute efficiently the stellar fidelities and the approximate stellar rank using Theorem~\ref{th:profiles}. 
For many copies, however, the stellar fidelities may be challenging to compute. To tackle this issue, we restrict the output to be a single copy of a target pure state, which is the case of interest in most resource conversion protocols, and we use Lemma~\ref{lem:upperbound} which gives bounds depending only on single-copy quantities to recover the case of pure states over few modes: 

\begin{theo}[Practical approximate conversion bounds for pure input]\label{th:conv-bound-approx-pure-single}
If a state $\delta$-close in trace distance to a single copy of $\ket{\bm\phi}$ can be obtained from $k$ copies of $\ket{\bm\psi}$ using a Gaussian protocol, then for all $0\le\epsilon\le1$ and all $n\in\mathbb N$,
\begin{equation}\label{eq:asr-conv-bound-approx-pure-single}
    \begin{cases}
        k\,r^\star_{\epsilon/k}(\bm\psi)\ge r^\star_{\epsilon+\delta}(\bm\phi),\\
        \delta\ge1-f^\star_{kn}(\bm\phi)-k(1-f_n^\star(\bm\psi)).
    \end{cases}
\end{equation}
Moreover, if a state $\delta$-close in trace distance to a single copy of $\ket{\bm\phi}$ can be obtained from $k$ copies of $\ket{\bm\psi}$ using a post-selected Gaussian protocol with any nonzero probability, then:
\begin{equation}\label{eq:asr-conv-bound-approximate0-pure-single}
    \begin{cases}
        k\,r^\star(\bm\psi)\ge r^\star_\delta(\bm\phi),\\
        \delta\ge1-f^\star_{k\,r^\star(\bm\psi)}(\bm\phi).
    \end{cases}
\end{equation}
\end{theo}

\noindent While these bounds already provide a practical means of assessing Gaussian conversion between pure states, many scenarios of interest involve mixed input states. In that case, computing stellar fidelities is challenging, but we may use pure stellar fidelities instead (see Eq.~(\ref{eq:puresfbound})). Together with sub-additive bounds from Lemma~\ref{lem:upperbound} we obtain:

\begin{theo}[Practical approximate conversion bounds for mixed input]\label{th:conv-bound-approx-mixed-single}
If a state $\delta$-close in trace distance to a single copy of $\ket{\bm\phi}$ can be obtained from $k$ copies of $\bm\rho$ using a Gaussian protocol, then for all $0\le\epsilon\le1$ and all $n\in\mathbb N$,
\begin{equation}\label{eq:asr-conv-bound-approx-mixed-single}
    \begin{cases}
        k\,r^\star_{\epsilon/k}(\bm\rho)\ge r^\star_{\epsilon+\sqrt\delta}(\bm\phi),\\
        \sqrt\delta\ge1-f^\star_{kn}(\bm\phi)-k(1-f_{n,\mathrm{pure}}^\star(\bm\rho)).
    \end{cases}
\end{equation}
Moreover, if a state $\delta$-close in trace distance to a single copy of $\ket{\bm\phi}$ can be obtained from $k$ copies of $\bm\rho$ using a post-selected Gaussian protocol with any nonzero probability, then:
\begin{equation}\label{eq:asr-conv-bound-approximate0-mixed-single}
    \begin{cases}
        k\,r^\star(\bm\rho)\ge r^\star_{\sqrt\delta}(\bm\phi),\\
        \sqrt\delta\ge1-f^\star_{k\,r^\star(\bm\rho)}(\bm\phi).
    \end{cases}
\end{equation}
\end{theo}

\noindent We give a proof of Theorems~\ref{th:conv-bound-approx-pure-single} and \ref{th:conv-bound-approx-mixed-single} in Appendix~\ref{app:single-copy}.

We explore the consequences of these bounds for assessing Gaussian conversion in the upcoming sections.

\subsection{No-go results for Gaussian conversion}
\label{sec:nogo}

The bounds in Theorem~\ref{th:asr-conv-bound-approx},  Theorem~\ref{th:conv-bound-approx-pure-single} and Theorem~\ref{th:conv-bound-approx-mixed-single} imply powerful no-goes for approximate Gaussian conversion, i.e., values of the trace distance error $\delta$ such that conversion is impossible. When using Gaussian protocols, a violation of any of the bounds in Eqs.~(\ref{eq:asr-conv-bound-approximate},\ref{eq:asr-conv-bound-approx-pure-single},\ref{eq:asr-conv-bound-approx-mixed-single}) rules out a possible conversion. When using post-selected Gaussian protocols, a violation of any of the bounds  in Eqs.~(\ref{eq:asr-conv-bound-approximate0},\ref{eq:asr-conv-bound-approximate0-pure-single},\ref{eq:asr-conv-bound-approximate0-mixed-single}) rules out a possible conversion with any nonzero probability.

These lead to several no-goes beyond state-of-the-art for Gaussian conversion between CV states. For instance, no-goes for Gaussian conversion have been obtained in~\cite{parellada2023no} in the case of deterministic and exact conversion restricted to passive Gaussian operations (linear optics). Several of these no-goes are subsumed by our results which are generally valid for approximate, deterministic and probabilistic Gaussian scenarios. We give a few examples hereafter and refer to the following section for a numerical analysis.

Consider the approximate, deterministic Gaussian conversion $\ket2\mapsto\ket1\otimes\ket1$. We have $f^\star_0(\ket1\otimes\ket1)\simeq0.25$ and $f^\star_0(\ket2)\simeq0.38$, so by Eq.~(\ref{eq:asr-conv-bound-approx-pure-single}) with $\ket{\bm\psi}=\ket1\otimes\ket1$, $\ket{\bm\phi}=\ket2$ and $k=1$,  approximate, deterministic Gaussian conversion from $\ket2$ to $\ket1\otimes\ket1$ cannot achieve a precision better than $0.13$ in trace distance. This is the vertical offset necessary to move the stellar profile of $\ket2$ below that of $\ket1\otimes\ket1$ (see Fig.~\ref{fig:profile1and11}).

On the other hand, Eq.~(\ref{eq:asr-conv-bound-approximate0-pure-single}) directly rules out exact, probabilistic Gaussian conversion from a state of finite stellar rank (such as any tensor product of finite superpositions of Fock states) to a state of infinite stellar rank (such as a cat state \cite{chabaud2020stellar} or a GKP state \cite{chabaud2021continuous}), no matter the success probability.
Moreover, it allows us to strengthen these no-go results to the case of approximate, probabilistic Gaussian conversion, as we show in the next section. Conversely, by setting a desired precision parameter $\delta$, Eq.~(\ref{eq:asr-conv-bound-approximate0-pure-single}) rules out approximate Gaussian conversion with any success probability using less than $\lceil r^\star_\delta(\bm\phi)/r^\star(\bm\psi)\rceil$ copies of the input state $\ket{\bm\psi}$.

Note that the (approximate) stellar rank may capture different non-Gaussian features than other monotones like WLN \cite{albarelli2018resource}. In particular, in the probabilistic setting, the bound in Eq.~(\ref{eq:asr-conv-bound-approximate0-pure-single}) becomes trivial if the input state has infinite stellar rank. In this case, it may still be beneficial to use WLN which can lead to nontrivial predictions, albeit in the exact conversion setting.


\section{Assessing approximate Gaussian conversion}
\label{sec:num}

It is often desirable to minimize the resources required to perform any quantum computational task or protocol.
Resource monotones are an important tool to aid this problem by offering the possibility to obtain fundamental lower bounds on how many resources are required for a given task. If a protocol saturates these bounds, there is no hope on improving the protocol further. However, if there is a gap between a lower bound and the best existing protocol there is the possibility of improving the protocol to use the resource more efficiently and improve experimental practicality. 

In this section, we provide a general methodology to harness our results on the approximate stellar rank in order to find lower bounds on Gaussian state conversion protocols, in the setting of both approximate and probabilistic conversions. We then illustrate this methodology on specific examples inspired by existing protocols. 

\subsection{Methodology}

In practice, we use Theorem~\ref{th:conv-bound-approx-pure-single} as our main tool for assessing Gaussian conversion protocols. To do so, we follow these steps:

\begin{enumerate}
    \item Choose a (pure state, many-to-one) Gaussian conversion scenario, given by $k$ copies of a pure input state $\ket{\bm\psi}$ and a target pure state $\ket{\bm\phi}$.
    \item Compute numerically the profiles of stellar fidelities for $\ket{\bm\psi}$ and $\ket{\bm\phi}$ using Theorem~\ref{th:profiles}.
    \item Compute the minimal allowed error in trace distance $\delta$ to the target state using Eq.~(\ref{eq:asr-conv-bound-approx-pure-single}) for Gaussian protocols and Eq.~(\ref{eq:asr-conv-bound-approximate0-pure-single}) for post-selected Gaussian protocols.
    \item Compare the approximation error of the Gaussian conversion protocol with the minimal allowed error.
\end{enumerate}

\noindent We build on top of the {\tt stellar-rank-numerics} library to assess Gaussian convertibility. For more details see Appendix~\ref{app:num}. All the code and scripts to generate the data and the figures presented in the following paragraphs are available online \cite{stellarnumerics}.

Notable Gaussian conversion protocols that were previously studied include: conversion of two single-photons into a two-photon state \cite{albarelli2018resource};  conversion of Fock states to cat states \cite{Ourjoumtsev2007generation};  synthesis of  GKP states from several non-Gaussian states \cite{Etesse:2014ty, Weigand2018generating, takase2024generation,pizzimenti2024optical}; conversion of tri-squeezed state to cubic phase state \cite{zheng2020gaussian, hahn2022deterministic}; $n$-photon added or $n$-photon subtracted squeezed states into cat states \cite{hahn2022deterministic};  binomial states into GKP states \cite{Zheng2023gaussian}; superposition of vacuum and single-photon state into arbitrary non-Gaussian states \cite{Etesse:14};  from photon-subtracted squeezed state into arbitrary non-Gaussian states \cite{Arzani2017polynomial};  from two copies of a cubic phase state to a cubic phase state with higher cubicity \cite{PhysRevA.97.062337}. 

In what follows, we focus on specific approximate Gaussian conversion scenarios inspired by the above list of protocols, and we investigate numerically both deterministic and probabilistic conversions.

\subsection{Deterministic Gaussian conversion between pure states}

We can employ Eq.~(\ref{eq:asr-conv-bound-approx-pure-single}) in Theorem~\ref{th:conv-bound-approx-pure-single} to compute lower bounds on the number of copies $k$ required to achieve a certain trace distance precision $\delta$, given a deterministic Gaussian conversion scenario $\ket{\bm\psi}^{\otimes k}\mapsto\ket{\bm\phi}$. We analyse two specific examples in what follows, focusing on the deterministic preparation of GKP states and cubic phase states.

\begin{figure}[t]
\centering
\includegraphics[width=\linewidth]{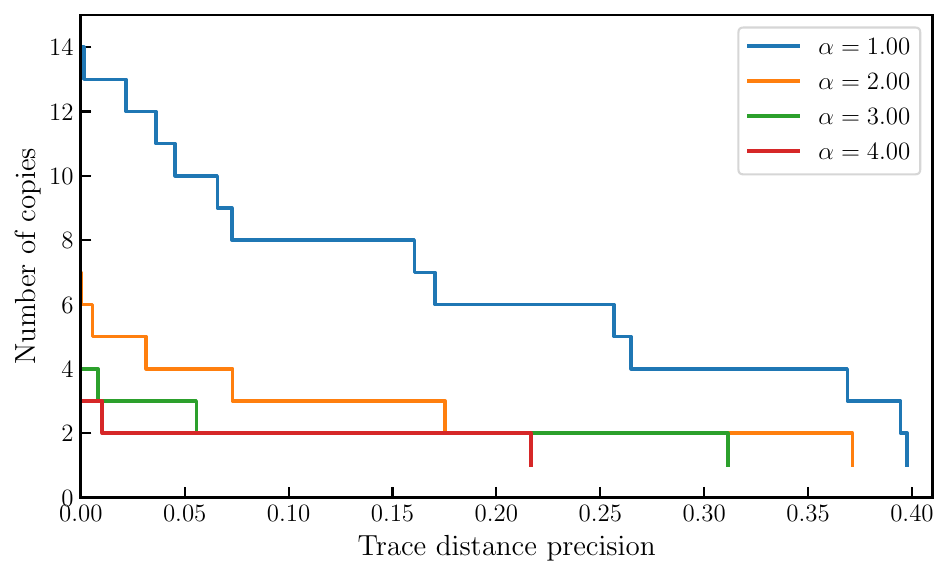}
\caption{Lower bounds on the number of copies of odd cat states required to achieve a given trace distance precision with a GKP state with $\Delta=0.3$, via deterministic Gaussian conversion, for various cat state amplitudes $\alpha=1,2,3,4$. All profiles used for this figure were computed up to a rank of 30 with 400 iterations in the optimisation except for $\alpha = 4$ where 500 iterations were used.}
\label{fig:cattogkp1}
\end{figure}

An important example of deterministic Gaussian conversions are cat breeding protocols where one takes multiple copies of a cat state and transforms them in a GKP state such as proposed in~\cite{Weigand2018generating}. Hereafter, we use the convention in~\cite[Approximation 1]{matsuura2020equivalence} for GKP states (the logical qubit state \(\ket{0}\)), with $\kappa=\Delta$. Using Theorem~\ref{th:conv-bound-approx-pure-single}, we obtain bounds on the number of copies of odd cat states $\propto\ket\alpha-\ket{-\alpha}$ required to generate a GKP state (logical 0 of a 2-dimensional codespace) with $\Delta=0.3$, for various cat state amplitudes $\alpha=1,2,3,4$. With these parameters, the  GKP state is approximated by a superposition of 9 Gaussian states. These bounds are displayed in Fig.~\ref{fig:cattogkp1}. For each amplitude $\alpha$, the plotted line delineates two regions in the parameter space $(\delta,k)$: below is an infeasibility region, where our bounds prove that deterministic Gaussian conversion is impossible; above is a possibility region, where our bounds cannot rule out the existence of a deterministic Gaussian conversion protocol with these parameters. Observe that, as the cat state amplitude is increased, so increases its non-Gaussianity, thereby shrinking the infeasibility region for converting to the 
target state.

Another example is the deterministic Gaussian conversion from a tri-squeezed state $e^{it(\hat a^3+\hat a^{\dag3})}\ket0$ to a cubic phase state $e^{ic\hat q^3}\hat S(\xi)\ket0$, discussed in~\cite{zheng2020gaussian,hahn2022deterministic}. Cubic phase states are important for implementing quantum computations based on CV systems as they allow to perform magic gates on the logical subspace of GKP codes~\cite{Gottesman2001}.
For the conversion between a tri-squeezed state with $t=0.125$ (resp.\ $t=0.15$) to a cubic phase state with squeezing of 5dB and $c=0.09$ (resp.\ $c=0.17$), a maximal fidelity of $F=0.9273$ (resp.\ $F=0.8557$) was reported in~\cite{zheng2020gaussian}.
Using that the trace distance may be expressed as
 $\sqrt{1-F}$ for pure states, we obtain a trace distance error of $\delta=0.2696$ and $\delta=0.3799$, respectively.
For single-copy to single-copy conversions, our corresponding bounds from Theorem~\ref{th:conv-bound-approx-pure-single} are computed as $\delta\ge0.01$ and $\delta\ge0.05$, respectively. This indicates that either stronger bounds may be derived for this conversion scenario, or better Gaussian protocols may be found. For more copies of tri-squeezed states, the bounds becomes trivial, suggesting that the sub-additivity property of the approximate stellar rank in Lemma~\ref{lem:upperbound} is generally not tight.

\subsection{Probabilistic Gaussian conversion between pure states}

We can also employ Eq.~(\ref{eq:asr-conv-bound-approximate0-pure-single}) in Theorem~\ref{th:conv-bound-approx-pure-single} to compute lower bounds on the number of copies $k$ required to achieve a certain trace distance precision $\delta$, given a probabilistic Gaussian conversion scenario $\ket{\bm\psi}^{\otimes k}\mapsto\ket{\bm\phi}$. In that case, note that the bound no longer depends on the specific input state $\ket{\bm\psi}$, but only on its stellar rank. We provide an example in what follows, when targeting the preparation of cat states.

\begin{figure}[t]
\centering
\includegraphics[width=\linewidth]{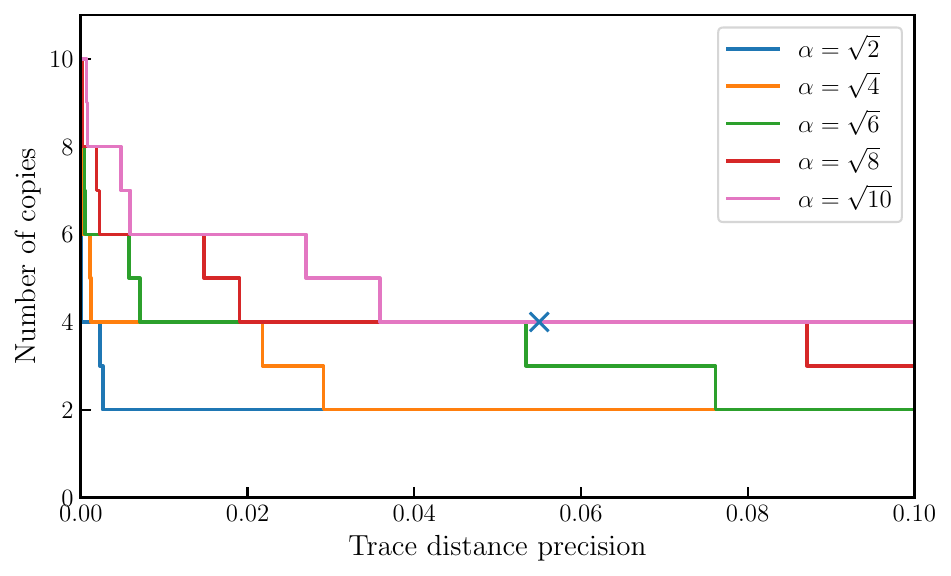}
\caption{Lower bounds on the number of copies of any state of stellar rank $1$  required to achieve a given trace distance precision with an even cat state via probabilistic Gaussian conversion, for various cat state amplitudes $\alpha \in \{\sqrt{2}, \sqrt{4}, \sqrt{6},\sqrt{8}, \sqrt{10}\}$. The blue cross depicts the performance of the protocol reported in~\cite[Figure 2b]{Etesse:2014ty}. All profiles used for this figure were computed up to a rank of 30 with 350 iterations in the optimisation.}
\label{fig:rank1tocat}
\end{figure}

In quantum optics, single-photon Fock states are a prototypical example of non-Gaussian resource states, as they form the basis of Boson Sampling \cite{Aaronson2013} and the Knill--Laflamme--Milburn scheme for universal quantum computing with linear optics \cite{knill2001scheme}. Moreover, Fock states can be combined using probabilistic Gaussian conversion to unlock more exotic non-Gaussian states. For instance, in~\cite{Etesse:2014ty}, a post-selected Gaussian protocol is introduced which probabilistically converts $k=2^n$ copies of a single-photon Fock state $\ket{1}$ to an even cat state $\propto\ket\alpha+\ket{-\alpha}$ with $\alpha=\sqrt n$. Using Theorem~\ref{th:conv-bound-approx-pure-single}, we obtain lower bounds on the number of copies of any state of stellar rank $1$ (including the Fock state $\ket{1}$) required to generate an even cat state, for various cat state amplitudes $\alpha=\sqrt n$ with $n=2,4,6,8,10$. These are displayed in Fig.~\ref{fig:rank1tocat}. As before, for each amplitude $\alpha$, the plotted line delineates two regions in the parameter space $(\delta,k)$: below is an infeasibility region, where our bounds prove that probabilistic Gaussian conversion is impossible, no matter the (nonzero) success probability; above is a possibility region, where our bounds cannot rule out the existence of a deterministic Gaussian conversion protocol with these parameters. Observe that, as the cat state amplitude is increased, so increases its non-Gaussianity, thereby increasing the infeasibility region as the target state gets more non-Gaussian.
For $k=4$ copies of a single-photon Fock state $\ket{1}$, targeting an even cat state of amplitude $\alpha=\sqrt2$, the fidelity of the probabilistic Gaussian protocol reported in~\cite[Figure 2b]{Etesse:2014ty} is $F=0.997$, which translates to a trace distance precision of $\delta=0.055$. For that precision parameter, our bounds indicate a minimal amount of $2$ copies required (see Fig.~\ref{fig:rank1tocat}). However, when targeting higher amplitudes with a similar precision, our bounds indicate that the protocol in~\cite{Etesse:2014ty} may be far from optimal, as it uses $2^n$ Fock states for engineering a cat state of amplitude $\alpha=\sqrt n$.

\subsection{Deterministic Gaussian conversion with mixed input states}

\begin{figure}[ht]
\centering
\includegraphics[width=\linewidth]{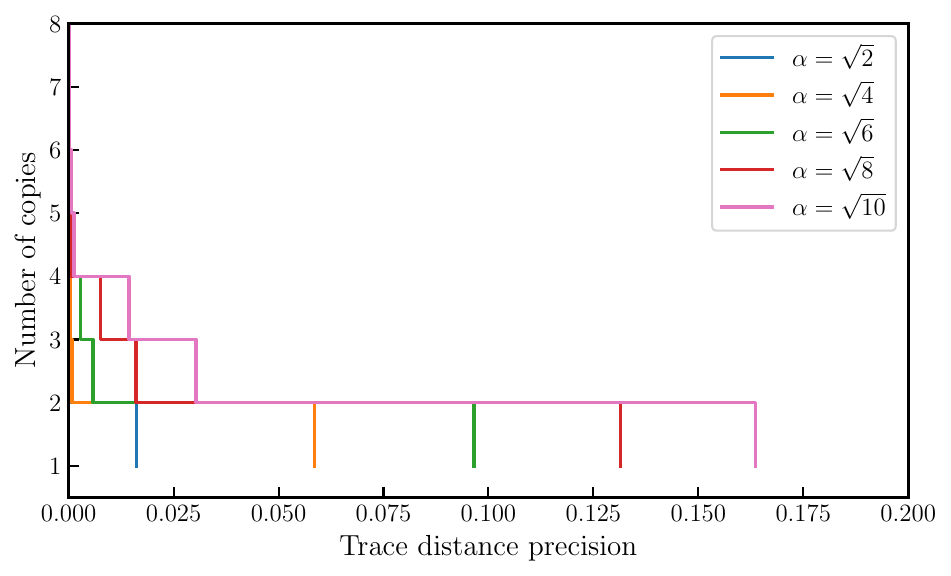}
\caption{Lower bounds on the number of copies of the mixed state \(\rho_p\) with \(p=0.2\)  required to achieve a given trace distance precision with an even cat state via deterministic Gaussian conversion, for various cat state amplitudes $\alpha \in \{\sqrt{2}, \sqrt{4}, \sqrt{6},\sqrt{8}, \sqrt{10}\}$. All profiles used for this figure were computed up to a rank of 30 with 250 (resp.\ 350) iterations in the optimisation of \(\rho_p\) (resp.\ cat states).}
\label{fig:mixedtocat}
\end{figure}

The previous examples focus on Gaussian conversion from pure states to pure states.
Following a similar methodology, we can employ instead Eq.~(\ref{eq:asr-conv-bound-approx-mixed-single}) in Theorem~\ref{th:conv-bound-approx-mixed-single} to compute lower bounds on the number of copies $k$ required to achieve a certain trace distance precision $\delta$, given a deterministic Gaussian conversion scenario from a mixed input state to an output pure state ${\bm\rho}^{\otimes k}\mapsto\ket{\bm\phi}$. We specifically focus on the case where the input state is \(\rho_p := p\ket0\!\bra0+(1-p)\ket1\!\bra1\) with \(p \in \left[0, 1\right]\) and the target state is an even cat of varying amplitude $\alpha=\sqrt n$ with $n=2,4,6,8,10$. In this case, since \(\rho_p\) is of stellar rank at most 1 by definition of the stellar rank for mixed states, only the $0^{\rm th}$ pure stellar fidelity \(f_{
0,\mathrm{pure}}^\star(\rho_p)\) defined in Eq.~(\ref{eq:puresf}) is required, and we show in Appendix~\ref{app:mixedproof} that it can be computed analytically. In Fig.~\ref{fig:mixedtocat} we show the lower bounds obtained numerically for \(p = 0.2\).

From a practical perspective, we expect the lower bounds on approximate convertibility with respect to the trace distance to be looser in the mixed-input case due to the fact that the trace distance $\delta$ from Theorem~\ref{th:conv-bound-approx-pure-single} is replaced by $\sqrt{\delta}$ in Theorem~\ref{th:conv-bound-approx-mixed-single}.


\section{Conclusion}
\label{sec:conclusion}

In this work, we have introduced the notion of approximate stellar rank, a robust and operational measure of non-Gaussianity for CV quantum states which generalises the stellar rank and that is easy to compute in most practical scenarios. We showed that this new measure is a non-Gaussian monotone and leads to a family of bounds for Gaussian conversion, both in the exact and approximate settings, including also probabilistic protocols. These bounds directly imply new no-go results for Gaussian conversion. Moreover, we showed how they can be used to assess the quality of Gaussian conversion protocols, by studying existing protocols.

Our results have broad applications for quantum state engineering in the CV setting. They imply that the non-Gaussian structure of CV quantum states can be revealed via profiles of stellar fidelities, including for states of infinite stellar rank. The formalism developed in this work allows us to assess Gaussian conversion in a very general setting. Moreover, our work opens up for several follow-up directions. These include benchmarking other existing Gaussian conversion protocols against the bounds stemming from the approximate stellar rank; generalising our framework to hybrid protocols where some gates or measurements are non-Gaussian, e.g., by bringing all non-Gaussianity to the input; finally, a further promising avenue for future research is generalising the computational interpretation of stellar rank~\cite{chabaud2023resources} to the approximate stellar rank, i.e., deriving an approximate classical simulation algorithm of bosonic quantum computations whose computational cost scales with the approximate stellar rank of the states involved in the computation.


\section*{Acknowledgements}

U.C.\ acknowledges interesting discussions with P.~Stornati, F.~Centrone, F.~Grosshans, S.~Mehraban, R.~Nehra, M.~Walschaers, H.~Yamasaki and R.~Takagi.
We thank S.F.E.~Oliviero and L.~Leone for pointing us to an error in an earlier version of this work. M.G. \ thanks T.~Martinez for critical code reviews and insightful discussions.
U.C., A.F.\ and G.F.\ acknowledge funding from the European Union’s Horizon Europe Framework Programme (EIC Pathfinder Challenge project Veriqub) under Grant Agreement No.\ 101114899.
G.F.\ acknowledges financial support from the Swedish Research Council (Vetenskapsrådet) through the project grant DAIQUIRI. G.F.\ and O.H.\ acknowledge support from the Knut and Alice Wallenberg Foundation through the Wallenberg Center for Quantum Technology (WACQT). O.H.\ acknowledges support from CREST Grant Number JPMJCR23I3, Japan. M.G.\ acknowledges funding by the Hybrid Quantum Initiative (HQI) supported by France 2030 under ANR grant ANR-22-PNCQ-0002.


\bibliographystyle{linksen}
\bibliography{biblio}


\onecolumn\newpage
\appendix


\begin{center}
    {\huge Appendix}
\end{center}


\section{Properties of the approximate stellar rank}
\label{app:asr-nGmon}

In this section, we prove the properties of the $\epsilon$-approximate stellar rank stated in the main text. 

\medskip

\noindent\textbf{Proof of Theorem~\ref{th:asr-nGmon}.} We show that (i) the $\epsilon$-approximate stellar rank vanishes for Gaussian states and (ii) is non-increasing under Gaussian maps and (iii) classical mixing. Moreover we show that  (iv) when $\epsilon=0$, the $\epsilon$-approximate stellar rank is a monotone under post-selection with respect to Gaussian measurements.

\medskip

\noindent (i) The $\epsilon$-approximate stellar rank vanishes for Gaussian states:

Given a mixture of Gaussian states $\bm\rho$, we have $r^\star_0(\bm\rho)=r^\star(\bm\rho)=0$ and since the $\epsilon$-approximate stellar rank is a non-increasing non-negative function this gives $r^\star_\epsilon(\bm\rho)=0$ for all $0\le\epsilon\le1$.

\medskip

\noindent (ii) The $\epsilon$-approximate stellar rank is non-increasing under Gaussian maps:

Fix $0\le\epsilon\le1$, a density operator $\bm\rho$ and a Gaussian completely positive trace-preserving map $\mathcal G$. We write $\bm\sigma=\mathcal G(\bm\rho)$. By definition of the $\epsilon$-approximate stellar rank, there exists a density operator $\bm\tau$ such that $F(\bm\rho,\bm\tau)\ge1-\epsilon$ and $r^\star(\bm\tau)=r^\star_\epsilon(\bm\rho)$. Then,
\begin{equation}
    F(\bm\sigma,\mathcal G(\bm\tau))=F(\mathcal G(\bm\rho),\mathcal G(\bm\tau))\ge F(\bm\rho,\bm\tau)\ge1-\epsilon,
\end{equation}
where we used the fact that the fidelity does not decrease under quantum operations.
Hence, 
\begin{equation}\label{eq:repssigma}
    r^\star_\epsilon(\bm\sigma)\le r^\star(\mathcal G(\bm\tau)),
\end{equation}
by definition of the $\epsilon$-approximate stellar rank. Moreover,
\begin{equation}
    r^\star(\mathcal G(\bm\tau))\le r^\star(\bm\tau)=r^\star_\epsilon(\bm\rho),
\end{equation}
where we used the fact that the stellar rank is non-increasing under Gaussian maps \cite{chabaud2021holomorphic}. Together with Eq.~(\ref{eq:repssigma}) this concludes the proof. Note that a similar proof shows that the $\epsilon$-smoothed non-Gaussianity of formation is also non-increasing under Gaussian maps.\qed

\medskip

\noindent (iii) The $\epsilon$-approximate stellar rank is non-increasing under classical mixing:

Let $0\le p\le1$, let $0\le\epsilon\le1$ and let $\bm\rho_1,\bm\rho_2$ be two states. We aim to show that
\begin{equation}
    r^\star_\epsilon(p\bm\rho_1+(1-p)\bm\rho_2)\le\max(r^\star_\epsilon(\bm\rho_1),r^\star_\epsilon(\bm\rho_2)).
\end{equation}
We assume that both states have finite $\epsilon$-approximate stellar rank  (otherwise the result is trivial). By definition of the approximate stellar rank, there exists $\bm\tau_1,\bm\tau_2$ such that $F(\bm\tau_1,\bm\rho_1)\ge1-\epsilon$, $F(\bm\tau_2,\bm\rho_2)\ge1-\epsilon$, $r^\star_\epsilon(\bm\rho_1)=r^\star(\bm\tau_1)$ and $r^\star_\epsilon(\bm\rho_2)=r^\star(\bm\tau_2)$. Now, the joint concavity of the fidelity \cite[Property 9.2.2]{wilde2013quantum} ensures that
\begin{align}
    F(p\bm\tau_1+(1-p)\bm\tau_2,p\bm\rho_1+(1-p)\bm\rho_2)&\ge pF(\bm\tau_1,\bm\rho_1)+(1-p)F(\bm\tau_2,\bm\rho_2)\\
    &\ge1-\epsilon,
\end{align}
so $r^\star_\epsilon(p\bm\rho_1+(1-p)\bm\rho_2)\le r^\star(p\bm\tau_1+(1-p)\bm\tau_2)$. Since the stellar rank is non-increasing under classical mixing \cite[Eq.~(95)]{chabaud2021holomorphic}, this implies
\begin{align}
    r^\star_\epsilon(p\bm\rho_1+(1-p)\bm\rho_2)&\le\max(r^\star(\bm\tau_1),r^\star(\bm\tau_2))\\
    &=\max(r^\star_\epsilon(\bm\rho_1),r^\star_\epsilon(\bm\rho_2)),
\end{align}
which concludes the proof.
\qed

\medskip

\noindent (iv) For $\epsilon=0$, the $\epsilon$-approximate stellar rank is non-increasing when post-selecting on Gaussian measurements of subsystems:

For $\epsilon=0$, the $\epsilon$-approximate stellar rank is equal to the stellar rank, and the stellar rank is non-increasing when post-selecting on Gaussian measurements of subsystems~\cite[Corollary 4]{chabaud2021holomorphic}.\qed

\medskip

\noindent This concludes the proof of Theorem~\ref{th:asr-nGmon}.\hfill$\blacksquare$

\medskip

\noindent Note that the stellar rank (and thus the $\epsilon$-approximate stellar rank) is not convex in general: for instance, the stellar rank of the state $\frac12\ket0\!\bra0+\frac12\ket1\!\bra1$ is equal to $1$, while the average of the individual stellar ranks is equal to 1/2.

\medskip

\noindent\textbf{Proof of Lemma~\ref{lem:upperbound}.} The lower bound
\begin{equation}
    \max_ir^\star_\epsilon(\bm\rho_i)\le r^\star_{\epsilon}\left(\bigotimes_{i=1}^k\bm\rho_i\right)\!,
\end{equation}
is a direct consequence of the fact that the $\epsilon$-approximate stellar rank is non-increasing under partial trace, which was proven above. For the upper bound, we first pick density operators $\bm\tau_1,\dots,\bm\tau_k$ such that 
$F(\bm\tau_i,\bm\rho_i)\ge1-\epsilon_i$ and $r^\star(\bm\rho_i)=r^\star_{\epsilon_i}(\bm\rho_i)$ for all $i\in\{1,\dots,k\}$. 
We have
\begin{equation}
    \begin{aligned}
        F\left(\bigotimes_{i=1}^k\bm\tau_i,\bigotimes_{i=1}^k\bm\rho_i\right)&=\prod_{i=1}^kF(\bm\tau_i,\bm\rho_i)\\
        &\ge\prod_{i=1}^k(1-\epsilon_i)\\
        &\ge1-(\epsilon_1+\dots+\epsilon_k),
    \end{aligned}
\end{equation}
using the Weierstrass product inequality \cite{mitrinovic1970analytic} in the last line. By definition of the $\epsilon$-approximate stellar rank we obtain:
\begin{equation}
    r^\star_{|\bm\epsilon|}\left(\bigotimes_{i=1}^K\bm\rho_i\right)\le r^\star\left(\bigotimes_{i=1}^k\bm\tau_i\right)\le\sum_{i=1}^kr^\star(\bm\tau_i)=\sum_{i=1}^kr^\star_{\epsilon_i}(\bm\rho_i).
\end{equation}
\hfill$\blacksquare$

\section{Equivalence between stellar fidelities and approximate stellar rank}
\label{app:equiv}

In this section, we prove Lemma~\ref{lem:asr-sF} from the main text.

\medskip

\noindent\textbf{Proof of Lemma~\ref{lem:asr-sF}.}
Let us start by proving the following: for all $n\in\mathbb N$ and all $0\le\epsilon\le1$,
\be\label{eq:implication1}
    r^\star_\epsilon(\bm\rho)\le n\Rightarrow f^\star_n(\bm\rho)\ge1-\epsilon.
\ee
Indeed, if the approximate stellar rank satisfies $r^\star_\epsilon(\bm\rho)\le n$, then there exists a state $\bm\tau$ such that $F(\bm\rho, \bm\tau)\ge 1- \epsilon$ and $r^\star(\bm\tau)\le n$, so
\be 
f_n^\star(\bm\rho)\ge F(\bm\rho,\bm\tau)\ge 1-\epsilon.
\ee 

Now by the definition of stellar fidelities, assuming $f^\star_n(\bm\rho)\ge1-\epsilon$ implies that for all $\epsilon'>\epsilon$, there exists a state $\bm\tau$ such that $F(\bm\rho, \bm\tau) > 1 - \epsilon'$ and $r^\star(\bm\tau) \le n$, which also implies that the $\epsilon'$-approximate stellar rank of $\bm\rho$ is at most $n$ for all $\epsilon'>\epsilon$. We thus obtain, for all $n\in\mathbb N$ and all $0\le\epsilon\le1$,
\begin{align}
    \label{eq:implinter1}f^\star_n(\bm\rho)\ge1-\epsilon&\Rightarrow\forall\epsilon'>\epsilon,\,r^\star_{\epsilon'}(\bm\rho)\le n\\
    \label{eq:implinter2}&\Rightarrow\forall\epsilon'>\epsilon,\,f^\star_n(\bm\rho)\ge1-\epsilon'\\
    \label{eq:implinter3}&\Rightarrow f^\star_n(\bm\rho)\ge1-\epsilon,
\end{align}
where we have used Eq.~(\ref{eq:implication1}) in the second line, and where we have taken the limit $\epsilon'\rightarrow\epsilon$ in the last line. This shows that all implications (\ref{eq:implinter1}-\ref{eq:implinter3}) are equivalences and thus, for all $n\in\mathbb N$ and all $0\le\epsilon\le1$,
\begin{equation}\label{eq:implication2}
     f^\star_n(\bm\rho)\ge1-\epsilon\Leftrightarrow\forall\epsilon'>\epsilon,\,r^\star_{\epsilon'}(\bm\rho)\le n.
\end{equation}

Now for all $n\in\mathbb N\setminus\{0\}$ and all $0\le\epsilon\le1$,
\begin{equation}\label{eq:implication3}
    \begin{aligned}
    \epsilon<1-f^\star_{n-1}(\bm\rho)&\Rightarrow r^\star_\epsilon(\bm\rho)> n-1\\
    &\Rightarrow r^\star_\epsilon(\bm\rho)\ge n,
    \end{aligned}
\end{equation}
where we have used (the contrapositive of) Eq.~(\ref{eq:implication1}) in the first line and the fact that the approximate stellar rank is integer-valued in the second line. On the other hand, for all $n\in\mathbb N$ and all $0\le\epsilon\le1$,
\begin{equation}\label{eq:implication4}
    \begin{aligned}
    \epsilon>1-f^\star_n(\bm\rho)&\Rightarrow\exists\epsilon''<\epsilon,\,f^\star_n(\bm\rho)\ge1-\epsilon''\\
    &\Rightarrow\exists\epsilon''<\epsilon,\,\forall\epsilon'>\epsilon'',\,r^\star_{\epsilon'}(\bm\rho)\le n\\
    &\Rightarrow r^\star_{\epsilon}(\bm\rho)\le n,
    \end{aligned}
\end{equation}
where in the first line one can choose any $\epsilon''\in(1-f^\star_n(\bm\rho),\epsilon)$, and where we used Eq.~(\ref{eq:implication2}) in the second line.
Combining Eq.~(\ref{eq:implication3}) and Eq.~(\ref{eq:implication4}) we obtain, for all $1\le n<r^\star(\bm\rho)$,
\begin{equation}
    r^\star_\epsilon(\bm\rho) =
    \begin{cases} 
    r^\star(\bm\rho) & \text{for } 
    \epsilon\in[0,1-f^\star_{r^\star(\bm\rho)-1}(\bm\rho)),\\
    n &\text{for } 
    \epsilon\in(1-f^\star_n(\bm\rho),1-f^\star_{n-1}(\bm\rho)),\\
    0 &\text{for }\epsilon\in(1-f^\star_0(\bm\rho),1].
    \end{cases}
\end{equation}

We now turn to the pure state case. In that case, the optimised function $\bm\tau\mapsto F(\bm\tau,\bm\psi)$ in the definition of stellar fidelities becomes linear, and states of stellar rank less or equal to $n$ can be written as convex mixtures of pure states of stellar rank less or equal to $n$, by definition of the stellar rank of mixed states. This implies that we can reduce the optimisation over the set of of pure states of stellar rank less or equal to $n$. without loss of generality, i.e., for all $n\in\mathbb N$,
\begin{equation}
    f_n^\star(\bm\psi)=\sup_{\bm\phi,r^\star(\bm\phi)\le n}F(\bm\psi,\bm\phi).
\end{equation}
Now for all $n\in\mathbb N$, the set of pure states of stellar rank less or equal to $n$ is a closed set for the trace norm \cite[Theorem 5]{chabaud2021holomorphic}, and for any state $\ket{\bm\psi}$ the function $\bm\tau\mapsto F(\bm\tau,\bm\psi)$ is continuous for the trace norm (see Lemma~\ref{lem:diffFid}), so
\begin{equation}
    f_n^\star(\bm\psi)=\max_{\bm\phi,r^\star(\bm\phi)\le n}F(\bm\psi,\bm\phi).
\end{equation}
In particular, assuming $f_n^\star(\bm\psi)\ge1-\epsilon$ implies that there exists a pure state $\ket{\bm\phi}$ such that $r^\star(\bm\phi)\le n$ and $F(\bm\psi,\bm\phi)=f_n^\star(\bm\psi)\ge1-\epsilon$, which also implies that the $\epsilon$-approximate stellar rank of $\ket{\bm\psi}$ is at most $n$. Together with Eq.~(\ref{eq:implication1}) this shows that 
\begin{equation}
    f^\star_n(\bm\psi)\ge1-\epsilon\Leftrightarrow r^\star_\epsilon(\bm\psi)\le n,
\end{equation}
for all $n\in\mathbb N$ and all $0\le\epsilon\le1$, which directly implies
\begin{equation}
    \!r^\star_\epsilon(\bm\psi) =\!
    \begin{cases} 
    r^\star(\bm\psi) & \!\!\!\!\text{for } 
    \epsilon\!\in\![0,1-f^\star_{r^\star(\bm\psi)-1}(\bm\psi)),\\
    n & \!\!\!\!\text{for } 
    \epsilon\!\in\![1-f^\star_n(\bm\psi),1-f^\star_{n-1}(\bm\psi)),\\
    0 & \!\!\!\!\text{for } \epsilon\!\in\![1-f^\star_0(\bm\psi),1].
    \end{cases}
\end{equation}

\noindent This concludes the proof of Lemma~\ref{lem:asr-sF}.\hfill$\blacksquare$


\section{Pure stellar fidelities}
\label{app:puresf}

We define pure stellar fidelities as follows:

\begin{defi}[Pure stellar fidelities]\label{def:puresF}
The pure stellar fidelities of a state $\bm\rho$ are defined as
\begin{equation}\label{eq:purestellarF}
    f_{n,\mathrm{pure}}^\star(\bm\rho):=\sup_{\bm\phi,r^\star(\bm\phi)\le n}F(\bm\rho,\bm\phi),
\end{equation}
for all $n\in\mathbb N$.
\end{defi}

Compared to stellar fidelities (see Definition~\ref{def:sF}), pure stellar fidelities involve a maximization over only \textit{pure} states $\bm\phi$ of bounded stellar rank. As a result, $f_{n,\mathrm{pure}}^\star(\bm\rho)\le f_n^\star(\bm\rho)$ for all $n\in\mathbb N$.

We now show the expression of pure stellar fidelities given in Eq.~(\ref{eq:puresf}) in the main text:

\begin{theo}\label{th:puresf}
Let $n\in\mathbb N$ and let $\bm\rho$ be a $m$-mode target mixed state. Then, the maximum achievable fidelity with the target state $\bm\rho$ using $m$-mode pure states of finite stellar rank less or equal to $n$ is given by
\begin{equation}\label{eq:puresfprofiles}
f_{n,\mathrm{pure}}^\star(\bm\rho)=\max_{\hat G}[\mathrm{max}\,\mathrm{eig}\,(\bm\Pi_n\hat G^\dag\bm\rho\hat G\bm\Pi_n)],
\end{equation}
where $\bm\Pi_n=\sum_{|\bm p|\le n}\ket{\bm p}\!\bra{\bm p}$ and where the largest eigenvalue is maximized over all $m$-mode Gaussian unitary operations of the form $\hat G=\hat U\hat S\hat D$, where $\hat S$ is a tensor product of squeezing operators with real-valued squeezing parameters, $\hat D$ is a tensor product of displacement operators and $\hat U$ is a passive linear operator.
\end{theo}

\noindent\textbf{Proof of Theorem~\ref{th:puresf}.} The proof follows the derivation in~\cite{fiuravsek2022efficient}:
\begin{equation}
    \begin{aligned}
        f_{n,\mathrm{pure}}^\star(\bm\rho)&=\sup_{\bm\phi,r^\star(\bm\phi)\le n}F(\bm\rho,\bm\phi)\\
        &=\sup_{\bm\phi,r^\star(\bm\phi)\le n}\bra{\bm\phi}\bm\rho\ket{\bm\phi}\\
        &=\sup_{\ket{\bm C},\hat G}\bra{\bm C}\hat G^\dag\bm\rho\hat G\ket{\bm C}\\
        &=\sup_{\ket{\bm C},\hat G}\bra{\bm C}\bm\Pi_n\hat G^\dag\bm\rho\hat G\bm\Pi_n\ket{\bm C}\\
        &=\max_{\hat G}[\mathrm{max}\,\mathrm{eig}\,(\bm\Pi_n\hat G^\dag\bm\rho\hat G\bm\Pi_n)],
    \end{aligned}
\end{equation}
where $\ket{\bm C}$ is a core state such that $\bm\Pi_n\ket{\bm C}=\ket{\bm C}$. Since any Gaussian unitary operation may be written as $\hat U\hat S\hat D\hat V$, where $\hat S$ is a tensor product of squeezing operators with real-valued squeezing parameters, $\hat D$ is a tensor product of displacement operators and $\hat U$ and $\hat V$ are passive linear operators, and given that $\hat V\Pi_n\hat V^\dag=\Pi_n$, the operation $\hat V$ can be omitted in the unitary $\hat G$, which completes the proof of Theorem~\ref{th:puresf}.

\hfill$\blacksquare$

In particular, Theorem~\ref{th:profiles} implies that for pure states we have $f_{n,\mathrm{pure}}^\star(\bm\psi)=f_n^\star(\bm\psi)$, i.e., an optimal bounded stellar rank approximation of a pure state is a pure state.

Moreover, in the specific case of $n=0$, the pure stellar fidelity simplifies to
\begin{equation}\label{eq:puresfprofiles0}
f_{0,\mathrm{pure}}^\star(\bm\rho)=\max_{\hat G}\bra{\bm0}\hat G^\dag\bm\rho\hat G\ket{\bm0}.
\end{equation}
%


\section{Exact Gaussian conversion bounds}
\label{app:sF-conv-bound-exact}

In this section, we prove Lemma~\ref{lem:asr-conv-bound-exact} from the main text.

\medskip

\noindent\textbf{Proof of Lemma~\ref{lem:asr-conv-bound-exact}.}
By Theorem~\ref{th:asr-nGmon}, for all $0\le\epsilon\le1$, the $\epsilon$-approximate stellar rank is a monotone under Gaussian protocols.
Hence, if $\bm\sigma$ can be obtained from $\bm\rho$ using a Gaussian protocol, then for all $0\le\epsilon\le1$,
\begin{equation}
    r^\star_\epsilon(\bm\rho)\ge r^\star_\epsilon(\bm\sigma). 
\end{equation}
In that case, for all $n\in\mathbb N$,
\begin{equation}\label{eq:implinter4}
    r^\star_\epsilon(\bm\rho)\le n\Rightarrow r^\star_\epsilon(\bm\sigma)\le n,
\end{equation}
for all $0\le\epsilon\le1$, and thus:
\begin{equation}
    \begin{aligned}
    \exists\epsilon''<\epsilon,\,f^\star_n(\bm\rho)\ge1-\epsilon''&\Rightarrow\exists\epsilon''<\epsilon,\,\forall\epsilon'>\epsilon'',\,r^\star_{\epsilon'}(\bm\rho)\ge n\\
    &\Rightarrow r^\star_\epsilon(\bm\rho)\ge n\\
    &\Rightarrow r^\star_\epsilon(\bm\sigma)\le n\\
    &\Rightarrow f_n^\star(\bm\sigma)\ge1-\epsilon,
    \end{aligned}
\end{equation}
where we have used Eq.~(\ref{eq:implication2}) in the first line, Eq.~(\ref{eq:implinter4}) in the third line, and Eq.~(\ref{eq:implication1}) in the last line. By contraposition, for all $0\le\epsilon\le1$ and all $n\in\mathbb N$,
\begin{equation}
    \begin{aligned}
    f_n^\star(\bm\sigma)<1-\epsilon&\Rightarrow\forall\epsilon''<\epsilon,\,f^\star_n(\bm\rho)<1-\epsilon''\\
    &\Rightarrow f^\star_n(\bm\rho)\le1-\epsilon,
    \end{aligned}
\end{equation}
where we have taken the limit $\epsilon''\rightarrow\epsilon$, so 
\begin{equation}
    f^\star_n(\bm\rho)\le f^\star_n(\bm\sigma),
\end{equation}
for all $n\in\mathbb N$.

Moreover, Theorem~\ref{th:asr-nGmon} also shows that, for $\epsilon=0$, the $\epsilon$-approximate stellar rank is a monotone under post-selected Gaussian protocols. Hence, if $\bm\sigma$ can be obtained from $\bm\rho$ using a post-selected Gaussian protocol with any nonzero probability, then
\begin{equation}
    r^\star(\bm\rho)\ge r^\star(\bm\sigma).
\end{equation}
This concludes the proof of Lemma~\ref{lem:asr-conv-bound-exact}.\hfill$\blacksquare$


\section{Approximate Gaussian conversion bounds}
\label{app:asr-conv-bound-approx}

In this section we prove Theorem~\ref{th:asr-conv-bound-approx} from the main text.

\medskip

\noindent We make use the following result: 

\begin{lem}\label{lem:diffFid}
For density operators $\bm\rho$, $\bm\sigma$ and $\bm\tau$, we have
\begin{equation}
    \left|F(\bm\rho,\bm\tau)-F(\bm\sigma,\bm\tau)\right|\le\sqrt{2D(\bm\rho,\bm\sigma)},
\end{equation}
where $D$ is the trace norm. Moreover, if $\bm\sigma=\bm\phi$ is a pure state,
\begin{equation}\label{eq:diffinter}
    \left|F(\bm\rho,\bm\tau)-F(\bm\phi,\bm\tau)\right|\le\sqrt{D(\bm\rho,\bm\sigma)}.
\end{equation}
Finally, if $\bm\rho=\bm\psi$ and $\bm\sigma=\bm\phi$ are both pure states,
\begin{equation}\label{eq:diffF}
    \left|F(\bm\psi,\bm\tau)-F(\bm\phi,\bm\tau)\right|\le D(\bm\psi,\bm\phi).
\end{equation}
\end{lem}

\noindent \textbf{Proof of Lemma~\ref{lem:diffFid}.} For density operators $\bm\rho$, $\bm\sigma$ and $\bm\tau$, we have \cite[Lemma 1]{rastegin2003lower}:
\begin{equation}
    |F(\bm\rho,\bm\tau)-F(\bm\sigma,\bm\tau)|\le\sqrt{1-F(\bm\rho,\bm\sigma)}.
\end{equation}
By the Fuchs--van de Graaf inequalities, $1-\sqrt F\le D$, so $\sqrt{1-F}\le\sqrt{1-(1-D)^2}\le\sqrt{2D}$, which concludes the proof for $\bm\rho$ and $\bm\sigma$ mixed states. 

When one of the states is pure, the Fuchs--van de Graaf inequality reads $1-F\le D$ so $\sqrt{1-F}\le\sqrt D$, which concludes the proof for $\bm\rho$ mixed state and $\bm\sigma=\bm\phi$ pure state.

For pure states $\bm\rho=\bm\psi$ and $\bm\sigma=\bm\phi$, we follow the proof from~\cite[Lemma 1]{chabaud2020building}: we consider the binary POVM $\{\bm\tau,\mathbb I-\bm\tau\}$ and we write $P_{\bm\psi}$ and $P_{\bm\phi}$ the corresponding probability distributions for the states $\bm\psi$ and $\bm\phi$, respectively. Then,
\begin{equation}
    \begin{aligned}
        \|P_{\bm\psi}-P_{\bm\phi}\|_\text{tvd}&=\frac12(|P_{\bm\psi}(0)-P_{\bm\phi}(0)|+|P_{\bm\psi}(1)-P_{\bm\phi}(1)|)\\
        &=|P_{\bm\psi}(0)-P_{\bm\phi}(0)|,
    \end{aligned}
\end{equation}
where $\|\cdot\|_\text{tvd}$ denotes the total variation distance, and
\begin{equation}
    \begin{aligned}
        |F(\bm\psi,\bm\tau)-F(\bm\phi,\bm\tau)|&=|\bra{\bm\psi}\bm\tau\ket{\bm\psi}-\bra{\bm\phi}\bm\tau\ket{\bm\phi}|\\
        &=|P_{\bm\psi}(0)-P_{\bm\phi}(0)|\\
        &=\|P_{\bm\psi}-P_{\bm\phi}\|_\text{tvd}\\
        &\le D(\bm\psi,\bm\phi),
    \end{aligned}
\end{equation}
by the variational definition of the trace distance.\qed

\medskip

\noindent\textbf{Proof of Theorem~\ref{th:asr-conv-bound-approx}.} Let $\bm\omega$ be a density operator such that $D(\bm\omega,\bm\sigma)\le\delta$ and $\bm\omega$ can be obtained from $\bm\rho$ using a Gaussian protocol.

Lemma~\ref{lem:diffFid} implies that the set $\{\bm\tau|F(\bm\sigma,\bm\tau)\ge1-(\epsilon+\sqrt{2\delta})\}$ contains the set $\{\bm\tau|F(\bm\omega,\bm\tau)\ge1-\epsilon\}$, so
\begin{equation}\label{eq:conv-bound-asr-inter}
    \begin{aligned}
        r^\star_{\epsilon+\sqrt{2\delta}}(\bm\sigma)&=\min_{\bm\tau,F(\bm\sigma,\bm\tau)\ge1-(\epsilon+\sqrt{2\delta})}r^\star(\bm\tau)\\
        &\le\min_{\bm\tau,F(\bm\omega,\bm\tau)\ge1-\epsilon}r^\star(\bm\tau)\\
        &=r^\star_\epsilon(\bm\omega).
    \end{aligned}
\end{equation}
With Lemma~\ref{lem:asr-conv-bound-exact} we have $r^\star_\epsilon(\bm\rho)\ge r^\star_\epsilon(\bm\omega)$, so we obtain
\begin{equation}
    r^\star_\epsilon(\bm\rho)\ge r^\star_{\epsilon+\sqrt{2\delta}}(\bm\sigma),
\end{equation}
for all $0\le\epsilon\le1$. The rest of the proof is analogous to that of Lemma~\ref{lem:asr-conv-bound-exact}: in that case, for all $n\in\mathbb N$,
\begin{equation}\label{eq:implinter4new}
    r^\star_\epsilon(\bm\rho)\le n\Rightarrow r^\star_{\epsilon+\sqrt{2\delta}}(\bm\sigma)\le n,
\end{equation}
for all $0\le\epsilon\le1$, and thus:
\begin{equation}
    \begin{aligned}
    \exists\epsilon''<\epsilon,\,f^\star_n(\bm\rho)\ge1-\epsilon''&\Rightarrow\exists\epsilon''<\epsilon,\,\forall\epsilon'>\epsilon'',\,r^\star_{\epsilon'}(\bm\rho)\ge n\\
    &\Rightarrow r^\star_\epsilon(\bm\rho)\ge n\\
    &\Rightarrow r^\star_{\epsilon+\sqrt{2\delta}}(\bm\sigma)\le n\\
    &\Rightarrow f_n^\star(\bm\sigma)\ge1-(\epsilon+\sqrt{2\delta}),
    \end{aligned}
\end{equation}
where we have used Eq.~(\ref{eq:implication2}) in the first line, Eq.~(\ref{eq:implinter4new}) in the third line, and Eq.~(\ref{eq:implication1}) in the last line. By contraposition, for all $0\le\epsilon\le1$ and all $n\in\mathbb N$,
\begin{equation}
    \begin{aligned}
    f_n^\star(\bm\sigma)+\sqrt{2\delta}<1-\epsilon&\Rightarrow\forall\epsilon''<\epsilon,\,f^\star_n(\bm\rho)<1-\epsilon''\\
    &\Rightarrow f^\star_n(\bm\rho)\le1-\epsilon,
    \end{aligned}
\end{equation}
where we have taken the limit $\epsilon''\rightarrow\epsilon$, so 
\begin{equation}
    f^\star_n(\bm\rho)\le f^\star_n(\bm\sigma)+\sqrt{2\delta},
\end{equation}
for all $n\in\mathbb N$.

Now if $\bm\omega$ is a density operator such that $D(\bm\omega,\bm\sigma)\le\delta$ and $\bm\omega$ can be obtained from $\bm\rho$ using a post-selected Gaussian protocol, the same proof with $\epsilon=0$ shows that
\begin{equation}
    r^\star(\bm\rho)\ge r^\star_{\sqrt{2\delta}}(\bm\sigma),
\end{equation}
for all $0\le\epsilon\le1$.

Finally, all proofs are identical when $\bm\sigma=\bm\phi$ is a pure state by replacing $\sqrt{2\delta}$ with $\sqrt\delta$, or when both $\bm\rho=\bm\psi$ and $\bm\sigma=\bm\phi$ are pure states by replacing $\sqrt{2\delta}$ with $\delta$. In the latter case, we obtain
\begin{equation}\label{eq:singlecopybound}
    r^\star_\epsilon(\bm\psi)\ge r^\star_{\epsilon+\delta}(\bm\phi),
\end{equation}
for all $0\le\epsilon\le1$ when a state $\delta$-close in trace distance to $\ket{\bm\phi}$ can be obtained from $\ket{\bm\psi}$ using a Gaussian protocol, while
\begin{equation}\label{eq:singlecopybound_ps}
    r^\star(\bm\psi)\ge r^\star_\delta(\bm\phi),
\end{equation}
when a state $\delta$-close in trace distance to $\ket{\bm\phi}$ can be obtained from $\ket{\bm\psi}$ using a post-selected Gaussian protocol.
This concludes the proof of Theorem~\ref{th:asr-conv-bound-approx}.\hfill$\blacksquare$

\section{Approximate Gaussian conversion bounds based on single-copy quantities}
\label{app:single-copy}

In this section, we prove Theorems~\ref{th:conv-bound-approx-pure-single} and \ref{th:conv-bound-approx-mixed-single} from the main text.

\medskip

\noindent\textbf{Proof of Theorem~\ref{th:conv-bound-approx-pure-single}.} If a state $\delta$-close in trace distance to a single copy of $\ket{\bm\phi}$ can be obtained from $k$ copies of $\ket{\bm\psi}$ using a Gaussian protocol, then by Eq.~(\ref{eq:singlecopybound}), for all $0\le\epsilon\le1$,
\begin{equation}
    r^\star_{\epsilon}(\bm\psi^{\otimes k})\ge r^\star_{\epsilon+\delta}(\bm\phi).
\end{equation}
Hence, the sub-additivity of the approximate stellar rank (see Lemma~\ref{lem:upperbound}) implies that
\begin{equation}
    k\,r^\star_{\epsilon/k}(\bm\psi)\ge r^\star_{\epsilon+\delta}(\bm\phi),
\end{equation}
for all $0\le\epsilon\le1$. In that case, for all $n\in\mathbb N$,
\begin{equation}\label{eq:implinter4new2}
    k\,r^\star_{\epsilon/k}(\bm\psi)\le kn\Rightarrow r^\star_{\epsilon+\delta}(\bm\phi)\le kn,
\end{equation}
for all $0\le\epsilon\le1$, and thus:
\begin{equation}
    \begin{aligned}
    \exists\epsilon''<\epsilon,\,f^\star_n(\bm\psi)\ge1-\epsilon''/k&\Rightarrow\exists\epsilon''<\epsilon,\,\forall\epsilon'>\epsilon'',\,r^\star_{\epsilon'/k}(\bm\psi)\ge n\\
    &\Rightarrow r^\star_{\epsilon/k}(\bm\psi)\ge n\\
    &\Rightarrow k\,r^\star_{\epsilon/k}(\bm\psi)\ge kn\\
    &\Rightarrow r^\star_{\epsilon+\delta}(\bm\phi)\le kn\\
    &\Rightarrow f_{kn}^\star(\bm\phi)\ge1-(\epsilon+\delta),
    \end{aligned}
\end{equation}
where we have used Eq.~(\ref{eq:implication2}) in the first line, Eq.~(\ref{eq:implinter4new2}) in the fourth line, and Eq.~(\ref{eq:implication1}) in the last line. By contraposition, for all $0\le\epsilon\le1$ and all $n\in\mathbb N$,
\begin{equation}
    \begin{aligned}
    f_{kn}^\star(\bm\phi)+\delta<1-\epsilon&\Rightarrow\forall\epsilon''<\epsilon,\,1-k(1-f^\star_n(\bm\psi))<1-\epsilon''\\
    &\Rightarrow 1-k(1-f^\star_n(\bm\psi))\le1-\epsilon,
    \end{aligned}
\end{equation}
where we have taken the limit $\epsilon''\rightarrow\epsilon$, so $1-k(1-f^\star_n(\bm\psi))\le f_{kn}^\star(\bm\phi)+\delta$ or equivalently
\begin{equation}\label{eq:finalstep}
    \delta\ge1-f_{kn}^\star(\bm\phi)-k(1-f^\star_n(\bm\psi)),
\end{equation}
for all $n\in\mathbb N$.

Now if a state $\delta$-close in trace distance to a single copy of $\ket{\bm\phi}$ can be obtained from $k$ copies of $\ket{\bm\psi}$ using a post-selected Gaussian protocol, then by Eq.~(\ref{eq:singlecopybound_ps}) we have $r^\star(\bm\psi^{\otimes k})\ge r^\star_\delta(\bm\phi)$, so
\begin{equation}\label{eq:finalstep2}
    k\,r^\star(\bm\psi)\ge r^\star_\delta(\bm\phi),
\end{equation}
since the stellar rank is additive for pure states \cite{chabaud2021holomorphic}. Finally, Eq.~(\ref{eq:implication1}) further implies
\begin{equation}
    \delta\ge1-f^\star_{k\,r^\star(\bm\psi)}(\bm\phi).
\end{equation}
\hfill$\blacksquare$

\noindent\textbf{Proof of Theorem~\ref{th:conv-bound-approx-mixed-single}.} The proof is identical to that of Theorem~\ref{th:conv-bound-approx-pure-single}, replacing $\bm\psi$ with $\bm\rho$ and $\delta$ by $\sqrt\delta$, and using $f_{n,\mathrm{pure}}^\star(\bm\rho)\le f_n^\star(\bm\rho)$ as an additional final step in Eq.~(\ref{eq:finalstep}), and the sub-additivity of stellar rank for mixed states \cite{chabaud2021holomorphic} for Eq.~(\ref{eq:finalstep2}).

\section{Details on the numerical implementation}
\label{app:num}

The basis of our numerical implementations is the Python {\tt stellar-rank-numerics} \cite{stellarnumerics} library which provides the necessary tools to compute single-mode stellar profiles. 

In the pure state case, stellar profiles are defined as \(\left\{f_n^\star(\psi)\right\}_{n\geq0 }\) with \(
f_n^\star(\psi)=\max_{\hat G}\bra{\psi}\hat G^\dag\Pi_n\hat G\ket{\psi},
\) as defined in Eq.~(\ref{eq:profiles}). For mixed states, we compute the lower bounds \(f_{
n,\mathrm{pure}}^\star(\rho)\) on \(f_n^\star(\rho)\) as described in Eqs.~(\ref{eq:puresf}) and (\ref{eq:puresfbound}). However, since we only examplify Gaussian conversion on the mixed state
\(\rho_p := p\ket0\!\bra0+(1-p)\ket1\!\bra1 \) which has stellar rank smaller or equal to one, the lower bound \(f_{
0,\mathrm{pure}}^\star(\rho)\) in Eq.~(\ref{eq:puresf}) reduces to the same expression as in the pure-state case Eq.~(\ref{eq:profiles}) with \(n=0\). Implementing the computation of \(f_{
n,\mathrm{pure}}^\star(\rho)\) for arbitrary mixed states using Eq.~(\ref{eq:puresf}) is left to future work.

As assessing (approximate) Gaussian conversion is an operation on a pair of stellar profiles, we add functionalities over the core of {\tt stellar-rank-numerics} to do so. 

As far as software development is concerned, we employ a simple and practical approach while ensuring openness, robustness, reproducibility, reusability and extensibility of the code. This software does not aim at being the most general, efficient or scalable. Indeed, its limits are reached for highly non-Gaussian states and multimode states are out of scope. However, we emphasise that \emph{(i)} it is enough for the usage at hand, \emph{(ii)} it builds a sound foundation for investigating other applications that rely on stellar profiles.

In what follows we give an overview of the design and the features of the software. We then discuss the convergence of the optimisation and provide insight showing that the numerical results obtained can be trusted. Finally, we quickly discuss how to improve the software in order to tackle further research problems.

\subsection{Structure and features}

\paragraph{Structure.}
Conceptually, the software is composed of:
\begin{itemize}
    \item an interface for specifying quantum CV states;
    \item tools to define objective functions, optimise them and compute stellar profiles;
    \item the definition of a {\tt StellarProfile} object along with utilities to save, load and plot;
    \item Gaussian conversion analysis tools operating on top of the {\tt StellarProfile} object and the core of the library.
\end{itemize}

\paragraph{Features.}

The software can express several type of states based on building blocks that are Fock states and Gaussian states. On the one hand, a statevector representation allows us to express any finite superposition of Fock states (e.g.\ binomial states). On the other hand, Gaussian states have their own interface and are parameterised by their displacement and squeezing parameters. This allows us to represent superpositions of Gaussian states as an iterable over a coefficient-Gaussian state pair. Some examples in this category of states are (multi-legged) cat states and approximate GKP states.

The library also allows to represent Hermitian operators by either providing the associated matrix in Fock space or by specifying a pure state decomposition. Density operators are thus expressible since they are Hermitian. However, no difference is made between general Hermitian operators and density matrices in the sense that there are currently no checks of positive-semidefiniteness or normalisation.

To numerically compute stellar profiles, recall that for a pure state, the stellar fidelity is
\begin{equation}
f_n^\star(\psi)=\max_{\hat G}\bra{\psi}\hat G^\dag\Pi_n\hat G\ket{\psi} = \max_{\hat G} \sum_{m \leq n} \left\vert\bra{m}{\hat G}\ket{\psi}\right\vert^2.
\end{equation}
Depending on the type of the input state, the objective function is computed differently:
\begin{itemize}
    \item if \(\ket{\psi}\) is expressed as a Gaussian state or a superposition of Gaussian states, we apply the Gaussian operator \(\hat G\) to every term in the decomposition, i.e.\ we update its displacement and squeezing parameters while keeping track of relative phases when needed;
    \item if \(\ket{\psi}\) is given as a statevector in the Fock basis, we expand it and compute the matrix elements of \(\bra{m}\hat G \ket{k}\) using the recursive method described in \cite{MQ2020}.
\end{itemize}
We then use a general-purpose global optimiser to finish the computation.

\medskip

As far as mixed states are concerned, the optimisation only currently supports those given as pure state decompositions when their stellar rank is at most one. In that case, we need only compute \(f_{
0,\mathrm{pure}}^\star(\rho)\) that reduces to the pure-state case Eq.~(\ref{eq:profiles}) with \(n=0\). This objective function being linear in the density matrix, we simply compute the objective function for each component of the decomposition and sum the results. Note that this procedure is not valid for mixed states of higher stellar rank. In general, in the derivation of Eq.~(\ref{eq:puresf}) (see Appendix~\ref{app:puresf}), we trade a linear objective function to be optimised over pure states of rank smaller or equal to \(n\) for a \emph{non-linear} function to be optimised over Gaussian unitaries.



For the sake of completeness, we provide a code example to compute a stellar profile using {\tt stellar-profile-numerics} in Listing~\ref{listing:usage}.

\begin{listing}[!ht]
\begin{minted}{python}
"""Script to produce stellar profiles."""

from stellar.cvstates import CatState
from stellar.params import Method, OptimisationParameters
from stellar.profile import compute_profile

# optimisation parameters 
# use the `gaussian` method since cat states are superpositions of Gaussian states
# fix the seed for reproducibility
max_rank = 10
niter = 350

params = OptimisationParameters(method=Method.gaussian, niter=niter, seed=652887) 

# Even cat of amplitude `amp` 
amp = 3
state = CatState(amplitude=amp, parity=False)

# compute the profile 
profile = compute_profile(ranks=list(range(max_rank + 1)), target_state=state, optim_params=params)

# save or plot...
profile.save_to_file(filename=...)
profile.draw(filename=...)
\end{minted}
\caption{Basic usage of {\tt stellar-rank-numerics} for an even cat state of amplitude 3.}
\label{listing:usage}
\end{listing}

\subsection{Convergence of stellar profiles}

Since the optimisation problem does not come with a certificate of convergence, it is important to build trust in the numerical results obtained. To do so, {\tt stellar-rank-numerics} includes several known and new results (both analytical and numerical) in its test suite:

\begin{itemize}
    \item all Gaussian states have stellar rank 0;
    \item Fock states \(\ket{n}\) have rank \(n\) and the stellar fidelities are checked against the known numerical results summarised in \cite{chabaud2020certification}. For \(n=1\), the analytical result is \(f_0^\star(\ket1)=\frac{3\sqrt3}{4e}\) \cite{chabaud2020stellar};
    \item comparison to the analytical computation of \(f_{0,\mathrm{pure}}^\star(\rho_p)\) for the mixed state \(\rho_p := p\ket0\!\bra0+(1-p)\ket1\!\bra1 \):

    \begin{align}
    \label{eq:mixed_state_analyt}
f_{0,\mathrm{pure}}^\star(\rho_p)&=\sup_{\hat G\in\mathcal G}\mathrm{eig}(\Pi_0\hat G\rho_p\hat G^\dag\Pi_0)\\
&=\begin{cases}\frac{1-p}eg_p(t^*)&p<\frac12,\\
p&p\ge\frac12,\end{cases}
    \end{align} where
\begin{itemize}
\item  $g_p(t)=\sqrt{(1+t)^3(1-t)}\,e^{\frac p{(1-p)(1+t)}}$,
\item  $t^*=\frac14(\sqrt{9 - 10 p/(1-p) + p^2/(1-p)^2}+t_\mathrm{min})$,
\item  $t_\mathrm{min}=\frac{2p-1}{1-p}$.
\end{itemize}
\end{itemize}
The detailed proof of this analytical expression is given in Appendix~\ref{app:mixedproof}.
In Fig.~\ref{fig:mixed_state_check}, we show the comparison of the analytical result Eq.~(\ref{eq:mixed_state_analyt}) to the numerical results and find perfect agreement.

\begin{figure}[h!]
    \centering
    \includegraphics[width=0.5\linewidth]{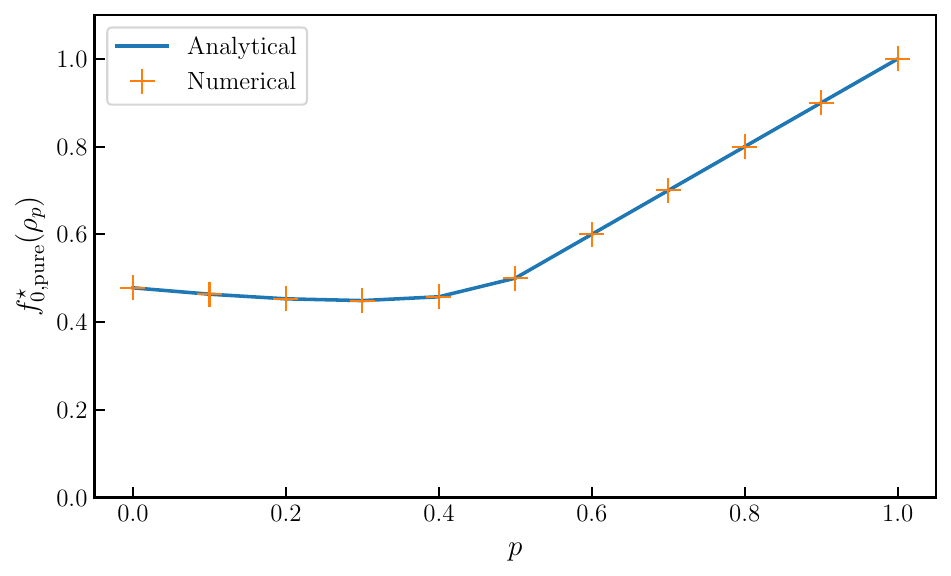}
    \caption{Comparison of the analytical and numerical stellar fidelity of \(f_{0,\mathrm{pure}}^\star(\rho_p)\) for several values of \(p\). The solid blue line shows the analytical result of Eq.~(\ref{eq:mixed_state_analyt}) while the orange crosses represent numerical computations using {\tt stellar-rank-numerics}. For all \(p\), the default optimisation parameters have been used.}
    \label{fig:mixed_state_check}
\end{figure}
These tests can be upgraded in the future using new analytical and numerical results to ensure correctness and non-regression of the library.

However, for highly non-Gaussian target states, finding the global optimum is more challenging, the final result of the optimization being dependent on the starting point of the optimisation and the random seed used. This indicates a complex optimisation landscapes and suggests using other optimisation methods, such as Riemannian optimization over the set of Gaussian unitaries~\cite{yao2024riemannian}.

\subsection{Conversion bounds}

In principle, the trace distance inequalities in Theorems~\ref{th:conv-bound-approx-pure-single} and~\ref{th:conv-bound-approx-mixed-single} have to hold for all \(n \in \mathbb N\). Numerically, it is of course not possible to check for all possible integers. However, we made sure to use numerical profiles so that for the highest ranks the stellar fidelities were close to 1. For higher numbers of copies where the range of computed $n$ is smaller, it might turn out that going to higher ranks might yield slightly stronger conversion bounds.

\subsection{Improvements and extensions}

The aim of our library is to showcase that it is possible to compute stellar profiles for a variety of single-mode states. If performance becomes an issue, there are several directions to be taken. The first one would be to parallelise the computation of the stellar profile since for each $n$ the computations are independent. The second is to optimise to computation of the state vector of Gaussian states in the Fock basis, since profiling shows this is where most of the time is spent. Also, a bigger improvement would be to investigate Riemannian optimisation techniques coupled with automatic differentiation \cite{yao2024riemannian,MrMustard}, especially in order to generalise our software to the multimode setting. All those improvements should be doable thanks to the modular nature of the software.


\section{Pure stellar fidelity of lossy single-photon states}
\label{app:mixedproof}
We consider the lossy single-photon state \(\rho_p := p\ket0\!\bra0+(1-p)\ket1\!\bra1 \), for $p\in[0,1]$. Since \(\rho_p\) is at most of stellar rank 1, its only non-trivial stellar fidelity corresponds to $n=0$. Hence, we only compute the pure stellar fidelity
\begin{align}
f_{0,\mathrm{pure}}^\star(\rho_p)&=\sup_{\hat G\in\mathcal G}\mathrm{eig}(\Pi_0\hat G\rho_p\hat G^\dag\Pi_0),
\end{align}
where \(\mathcal G\) is the group of single-mode Gaussian unitaries.
Writing $\hat G=\hat D(\gamma)\hat S(\xi)$, we have $\left\vert \bra 0\hat G\ket0\right\vert^2=\left\vert\langle-\gamma\vert\xi\rangle\right\vert^2$ and $\left\vert\bra 0\hat G\ket1\right\vert^2=\frac{|\gamma|^2}{\cosh^2r}\left\vert\langle-\gamma\vert\xi\rangle\right\vert^2$, with $\left\vert\langle-\gamma\vert\xi\rangle\right\vert^2=\frac1{\cosh r}e^{-|\gamma|^2-\frac12\tanh r(\gamma^2+\gamma^{*2})}$ (without loss of generality $\xi=r\ge0$). Hence

\begin{align}
f_{0,\mathrm{pure}}^\star(\rho_p)&=\sup_{\hat G\in\mathcal G}p\left\vert\bra 0\hat G\ket0\right\vert ^2+(1-p)\left\vert\bra 0\hat G\ket1\right\vert^2\\
&=\sup_{r,\gamma}\left[\frac1{\cosh r}e^{-|\gamma|^2-\frac12\tanh r(\gamma^2+\gamma^{*2})}\left(p+(1-p)\frac{|\gamma|^2}{\cosh^2r}\right)\right]\\
&=\sup_{r,x,y}\left[\frac1{\cosh r}e^{-(1+\tanh r)x^2-(1-\tanh r)y^2}\left(p+(1-p)\frac{(x^2+y^2)}{\cosh^2r}\right)\right]\\
&=\sup_{r,x,y}\left[\frac1{\cosh r}e^{-(1-\tanh r)(x^2+y^2)}e^{-(2\tanh r)x^2}\left(p+(1-p)\frac{(x^2+y^2)}{\cosh^2r}\right)\right]\\
&=\sup_{r,R}\left[\frac1{\cosh r}e^{-(1-\tanh r)R}\left(p+(1-p)\frac{R}{\cosh^2r}\right)\right]\quad(\text{setting }x=0).
\end{align} 
Here we have set $\gamma=x+iy$ and $R=x^2+y^2$. The best value for $R$ is $R^*=\frac1{1-\tanh r}-\frac{p(\cosh^2r)}{1-p}$ (or $R^*=0$ when this value is negative, i.e.\ for $1-p\le p(1-\tanh r)\cosh^2r$, in which case the supremum is $p$, when $p\ge\frac12$). We thus obtain (setting $t=\tanh r$)

\begin{align}
f_{0,\mathrm{pure}}^\star(\rho_p)&=\max\left\{p,\sup_{r;1-p>p(1-\tanh r)\cosh^2r}\left[\frac1{\cosh r}e^{-(1-\tanh r)R^*}\left(p+(1-p)\frac{R^*}{\cosh^2r}\right)\right]\right\}\\
&=\max\left\{p,\sup_{t\in[0,1],t>\frac{2p-1}{1-p}}\left[\frac{1-p}e\sqrt{(1+t)^3(1-t)}\,e^{\frac p{(1-p)(1+t)}}\right]\right\}.
\end{align} 

We define $g_p(t):=\sqrt{(1+t)^3(1-t)}\,e^{\frac p{(1-p)(1+t)}}$.
For $p\ge\frac12$, $g_p$ is non-increasing, so the max is reached at the smallest possible value $t_\mathrm{min}=\frac{2p-1}{1-p}\ge0$. The corresponding value $\frac{1-p}eg_p(t_\mathrm{min})$ gives $\frac12$ for $p=\frac12$ and is decreasing as a function of $p$, so in that case $f_{0,\mathrm{pure}}^\star(\rho_p)=p$,
For $p<\frac12$, we have $t_\mathrm{min}=\frac{2p-1}{1-p}<0$ and the max of $g_p$ is reached at 
    \begin{align}
        t^*~=\frac14(\sqrt{9 - 10 p/(1-p) + p^2/(1-p)^2}+t_\mathrm{min}).
    \end{align}
Hence:
\begin{align}
f_{0,\mathrm{pure}}^\star(\rho_p)=
\begin{cases}
\frac{1-p}eg_p(t^*)&p<\frac12,\\
p&p\ge\frac12,
\end{cases}
\end{align}

where:
\begin{itemize}
    \item $g_p(t)=\sqrt{(1+t)^3(1-t)}\,e^{\frac p{(1-p)(1+t)}}$,
    \item $t^*=\frac14(\sqrt{9 - 10 p/(1-p) + p^2/(1-p)^2}+t_\mathrm{min})$,
    \item $t_\mathrm{min}=\frac{2p-1}{1-p}$.
\end{itemize}

\noindent This concludes the proof.\hfill$\blacksquare$

\end{document}